\documentclass[preprint]{imsart}

\RequirePackage{amsthm,amsmath}
\RequirePackage[round]{natbib}
\RequirePackage[colorlinks,citecolor=blue,urlcolor=blue]{hyperref}
\usepackage{amssymb}
\usepackage{graphicx}
\usepackage{amsfonts}
\usepackage{amsbsy}
\usepackage{mathrsfs}
\usepackage{lscape}
\usepackage{float}
\theoremstyle{plain}

\usepackage{multirow}
\usepackage{tablefootnote}
\usepackage{threeparttable}
\usepackage{pdflscape}
\usepackage{enumerate}
\usepackage{pdfpages}
\usepackage{soul}
\usepackage{verbatim}
\usepackage{algorithm}
\usepackage{algpseudocode}

\startlocaldefs
\numberwithin{equation}{section}
\theoremstyle{plain}

\usepackage[english]{babel}

\endlocaldefs

\newcommand{\bB}{\mathbf{B}}
\newcommand{\bx}{\mathbf{x}}

\newcommand{\bS}{\mathbf{s}}

\newcommand{\bz}{\mathbf{z}}
\newcommand{\br}{\mathbf{r}}
\newcommand{\bM}{\mathbf{m}}
\newcommand{\bL}{\mathbf{L}}
\newcommand{\bI}{\mathbf{I}}
\newcommand{\bF}{\mathbf{F}}
\newcommand{\bu}{\mathbf{u}}
\newcommand{\bv}{\mathbf{v}}
\newcommand{\bT}{\mathbf{T}}
\newcommand{\bW}{\mathbf{W}}
\newcommand{\bU}{\mathbf{U}}
\newcommand{\bbeta}{\boldsymbol{\beta}}

\newcommand{\bTheta}{\boldsymbol{\Theta}}
\newcommand{\blambda}{\boldsymbol{\lambda}}
\newcommand{\bLambda}{\boldsymbol{\Lambda}}
\newcommand{\bSigma}{\boldsymbol{\Sigma}}
\newcommand{\bsigma}{\boldsymbol{\sigma}}
\newcommand{\brho}{\boldsymbol{\rho}}
\newcommand{\bEta}{\boldsymbol{\eta}}
\newcommand{\bmu}{\boldsymbol{\mu}}
\newcommand{\bDelta}{\boldsymbol{\Delta}}

\newcommand{\EE}{\mathbb{E}}

\DeclareMathOperator*{\argmax}{arg\,max}

\usepackage{xcolor}

\begin{document}

\begin{frontmatter}

\title{A Bayesian factor analysis model for high-dimensional microbiome count data}
\runtitle{Bayesian factor analysis technique for microbiome count data}

\begin{aug}
\author[A]{\fnms{Isma\"ila} \snm{Ba}\ead[label=e1]{ismaila.ba@umanitoba.ca}}
\author[A]{\fnms{Maxime} \snm{Turgeon}}
\author[C]{\fnms{Simona} \snm{Veniamin}}
\author[D]{\fnms{Juan} \snm{Jovel}}
\author[E]{\fnms{Richard} \snm{Miller}}
\author[E,F]{\fnms{Morag} \snm{Graham}}
\author[F]{\fnms{Christine} \snm{Bonner}}
\author[E]{\fnms{Charles N.} \snm{Bernstein}}
\author[G]{\fnms{Douglas L.} \snm{Arnold}}
\author[H]{\fnms{Amit} \snm{Bar-Or}}
\author[E]{\fnms{Ruth Ann} \snm{Marrie}}
\author[I]{\fnms{Julia} \snm{O'Mahony}}
\author[J]{\fnms{E. Ann} \snm{Yeh}}
\author[H]{\fnms{Brenda} \snm{Banwell}}
\author[K]{\fnms{Emmanuelle} \snm{Waubant}}
\author[E,F]{\fnms{Natalie} \snm{Knox}}
\author[E,F]{\fnms{Gary Van} \snm{Domselaar}}
\author[L]{\fnms{Ali I.} \snm{Mirza}}
\author[E,C]{\fnms{Heather} \snm{Armstrong}}
\author[A]{\fnms{Saman} \snm{Muthukumarana}}
\and
\author[B]{\fnms{Kevin} \snm{McGregor}}



\address[A]{Department of Statistics,
University of Manitoba, Canada \\
\printead{e1}}

\address[B]{Department of Mathematics and Statistics,
York University, Canada}

\address[C]{University of Alberta, Canada}

\address[D]{University of Calgary, Canada}

\address[E]{Department of Internal Medicine,
University of Manitoba, Canada}

\address[F]{Public Health Agency of Canada,
Winnipeg, Canada}

\address[G]{McGill University, Canada}

\address[H]{Perelman School of Medicine,
University of Pennsylvania, USA}

\address[I]{Mellen Center for Multiple Sclerosis,
Cleveland Clinic, USA}

\address[J]{Department of Pediatrics,
University of Toronto, Canada}

\address[K]{Weill Institute for Neurosciences,
University of California San Francisco, USA}

\address[L]{University of British Columbia, Canada}


\runauthor{Ba et al.}

\end{aug}

\begin{abstract}
\; Dimension reduction techniques are among the most essential analytical tools in the analysis of high-dimensional data. Generalized principal component analysis (PCA) is an extension to standard PCA that has been widely used to identify low-dimensional features in high-dimensional discrete data, such as binary, multi-category and count data. For microbiome count data in particular, the multinomial PCA is a natural counterpart of the standard PCA. However, this technique fails to account for the excessive number of zero values, which is frequently observed in microbiome count data. To allow for sparsity, zero-inflated multivariate distributions can be used. We propose a zero-inflated probabilistic PCA model for latent factor analysis. The proposed model is a fully Bayesian factor analysis technique that is appropriate for microbiome count data analysis. In addition, we use the mean-field-type variational family to approximate the marginal likelihood and develop a classification variational approximation algorithm to fit the model. We demonstrate the efficiency of our procedure for predictions based on the latent factors and the model parameters through simulation experiments, showcasing its superiority over competing methods. This efficiency is further illustrated with two real microbiome count datasets. The method is implemented in \texttt{R}.
\end{abstract}




\end{frontmatter}

\section{Introduction}

Dimension reduction techniques are nowadays needed in many fields, and the analysis of high-dimensional microbiome sequencing data is no exception. However, in microbiome studies, most existing reduction methods do not completely address the characteristics of the data such as discrete and abundance of zero counts. Other important characteristics of such data include for instance over-dispersion and complex dependence patterns among microbial taxa. Generalized principal component analysis, which is an extension of standard principal component analysis (PCA) to various types of discrete data from binary and multi-category responses to counts, has been widely used to identify low-dimensional features in high-dimensional data. For count data coming from microbiome studies in particular, the multinomial PCA (multivariate Poisson distribution) is a natural counterpart of the standard PCA (multivariate normal distribution). Due to the excessive number of zero values (also called extreme sparsity) frequently observed in microbiome sequencing data, this technique fails to appropriately account for the distribution of such data. Note that the zero values can be attributed to both biological zeros and technical zeros. More precisely, the biological zeros are zero values that genuinely represent the absence of a particular microbial taxon in a sample while the technical zeros are zero values that result from technical limitations, errors, or artifacts in the sequencing process. 

Zero-inflated multivariate distributions can be used to overcome the extreme sparsity in the data. For example, \cite{tang2019zero} proposed a zero-inflated generalized Dirichlet multinomial model to link covariates with taxonomic counts. It is worth pointing out that the Dirichlet multinomial model~\citep{mosimann1962compound,chen2013variable} assumes a Dirichlet distribution for modeling the taxa composition, which helps address over-dispersion, often observed in microbiome studies. Moreover, Dirichlet variates are always negatively correlated. The latter point indicates that the Dirichlet multinomial fails to fully represent the nature of microbiome data, which may allow positive associations between microbial communities~\citep{weiss2016correlation}. A natural and straightforward approach for overcoming this issue is to generalize the Dirichlet multinomial in a way that the resulting multinomial distribution can account for both positive and negative associations between microbes. This is the case of the generalized Dirichlet multinomial; however, the possibility of its variates to be positively correlated is relatively narrow~\citep{wang2017dirichlet}. In contrast to the Dirichlet distribution, the logistic normal (LN) distribution flexibly accommodates a large panel of dependency structure. Therefore, a model based on this distribution is preferable for its flexibility to allow for a more general correlation structure among taxon counts. The latter point motivates the logistic normal multinomial (LNM) model~\citep{aitchison1982statistical,billheimer2001statistical}. 

In a variable selection framework, \cite{xia2013logistic} considered the LNM model to associate microbial counts with covariates. They used respectively the multinomial and the logistic normal distributions to model the multivariate count data and the underlying microbial compositions. In a full Bayesian approach, \cite{billheimer2001statistical} used Monte Carlo Markov Chain (MCMC) to fit such an LNM model when the number of taxa is small. Variable selection was the main goal of the study in \cite{xia2013logistic}, so it is highly reasonable to assume that the number of taxa therein is moderate or large. Since the likelihood function for the LNM model is intractable, standard methods in parametric estimation cannot be directly applied. As an alternative, \cite{xia2013logistic} developed a Monte Carlo EM (MCEM) algorithm to overcome this difficulty. However, this method is more challenging from a computational point of view when hundreds or thousands of microbial taxa are considered, which is typical of count data collected from today’s high-throughput sequencing technologies. To address this computational challenge in high dimensions, \cite{zhang2019scalable} developed a scalable algorithm based on a stochastic approximation EM (SAEM) algorithm~\citep{delyon1999convergence} and Hamilton Monte Carlo (HMC) sampling~\citep{duane1987hybrid,neal2011mcmc}. In addition, \cite{zhang2019scalable} made the assumption to treat zero counts in multinomial data as sampling zeros. As pointed out by \cite{billheimer2001statistical}, this assumption may constitute a severe limitation in applications where one or more taxa are known to be absent, or where inference of absence is of interest. More recently, \cite{zeng2022zero} proposed a zero-inflated probabilistic PCA (ZIPPCA) model, which can be viewed as a low-rank version of the LNM model on the one hand, and as an extension of the probabilistic PCA~\citep{tipping1999probabilistic} on the other hand. They proposed an empirical Bayes approach for inferring microbial compositions. Moreover, they developed an iterative algorithm based on a variational approximation (VA)~\citep{ormerod2010explaining,blei2017variational} and classification for clusters~\citep{celeux1992classification} to a slightly lesser extent, to carry out the maximum likelihood estimation. They also provided theoretical results; that is, they proved the consistency and asymptotic normality of their VA estimator in a setting where the number of samples $n$ tends to infinity and the number of taxa $p$ is fixed. 

In this paper, we propose a zero-inflated probabilistic PCA (ZIPPCA) model for inferring microbial compositions. More specifically, we introduce a factor analysis technique with a complete Bayesian treatment. Note that the latter technique allows latent factor analysis that is appropriate for microbiome count data analysis. In addition, our proposed model is a full Bayesian version of the zero-inflated model introduced by \cite{zeng2022zero}, which makes the inference problems therein more complex. By taking advantage of the similarity between our model and that introduced in \cite{zeng2022zero}, we develop an efficient iterative algorithm based on mean-field variational Bayes (MFVB) approximation to find the maximum likelihood estimates of model parameters.

The remainder of the paper is structured as follows. In Section 2, we introduce the ZIPPCA model. We derive the variational lower bound and discuss the methodology as well as its implementation. In Section 3, we present the simulation study. Section 4 is devoted to the application of the proposed model to real microbiome datasets. A discussion of the results and a conclusion follow in Section 5.

\section{Methodology}\label{sec2}

\subsection{A Zero-Inflated Probabilistic PCA model}
Let us assume that we have a microbiome dataset with $n$ samples and $p$ bacterial taxa. We denote by $x_{ij}$ the observed read count of taxon $j$ in sample $i$. For the ith sample, the total count of all taxa, denoted by $M_i$, is determined by the sequencing depth, that is $M_i = \sum_{j=1}^p x_{ij}$. It is natural to model the stratified count data over p taxa, $\bx_i =(x_{i1},x_{i2},\ldots,x_{in})^\top$, as a multinomial distribution with index $M_i$ and vectors of probabilities $\brho_i = (\rho_{i1},\rho_{i2},\ldots,\rho_{ip})^\top$ where $0 < \rho_{ij} < 1$ for all $i=1,2,\ldots,n$ and $j=1,2,\ldots,p$, and $\sum_{j=1}^p \rho_{ij}=1$. Hence, the unknown taxon composition matrix, $\brho=(\rho_{ij}) \in \mathbb{R}^{n \times p}$, is an element of the $(p-1)$-dimensional simplex $\mathcal{S}^{p-1}$. It is worth mentioning that relative abundances or raw proportions are (naive) estimates of the compositions $\rho_{ij}$ obtained using the method of maximum likelihood. Other choices of the estimates of microbial compositions can be considered to better address issues encountered when analyzing microbiome sequencing data. For instance, sparsity (note that raw proportions often contain a large number of zeros) and over-dispersion are among one of them. To overcome these issues, particularly over-dispersion, a popular choice is often to assume that $\brho_i$ is random on the one hand and to model the taxon composition vector $\brho_i$ as a logistic normal (LN) distribution \citep{aitchison1982statistical} on the other hand. Note that the latter point relays on the additive log-ratio transform, which we annotate here by $\phi$. This transformation consists of taking the observations from $\mathcal{S}^{p-1}$ to $\mathcal{R}^{p-1}$, the $(p-1)$-dimensional Euclidean space. That is, for $\bmu=(\mu_1,\mu_2,\ldots,\mu_{p-1})^\top \in \mathcal{R}^{p-1}$ we have

\begin{align*}
\bmu = \phi(\brho) = \left\{ \log \left(\frac{\rho_1}{\rho_p}\right), \log \left(\frac{\rho_2}{\rho_p}\right), \ldots, \log \left(\frac{\rho_{p-1}}{\rho_p}\right) \right\}^\top.
\end{align*}
Since the transformation $\phi$ is bijective, its inverse denoted by $\phi^{-1}$ has the form
\begin{align*}
\brho = \phi^{-1}(\bmu) = \left\{  \frac{\exp(\mu_1)}{\sum_{j=1}^{p-1} \exp(\mu_{j})+1},\ldots,\frac{\exp(\mu_{p-1})}{\sum_{j=1}^{p-1} \exp(\mu_{j})+1}, \frac{1}{\sum_{j=1}^{p-1} \exp(\mu_{j})+1} \right\}^\top.
\end{align*}
In this setting, in order to induce a LN distribution on $\mathcal{S}^{p-1}$ i.e. $\brho_i$ follows a LN distribution, we have to assume multivariate normality of the transformed data $\bmu$. In addition, \cite{billheimer2001statistical} and more recently \cite{zeng2022zero} have used the LN distribution to model microbial compositions. Note that these two papers have used this assumption in two different frameworks: \cite{billheimer2001statistical} have considered a fully Bayesian procedure by introducing a prior over the mean and the covariance matrix in the LN distribution while \cite{zeng2022zero} have proposed an empirical Bayes approach (i.e. a comprise between the frequentist method and the fully Bayesian procedure) by assuming that the microbial composition $\brho_i$ follows a LN distribution but the hyperparameters are set to specific values.       

In this paper, we propose a fully Bayesian version of the zero-inflated model introduced by \cite{zeng2022zero} in order to make inferences about microbial compositions and to conduct factor analysis. More specifically, we consider a zero-inflated probabilistic PCA model of the form

\begin{flalign*}
& z_{ij} \overset{ind}{\sim} Bern(\eta_j), \\
&\boldsymbol{f}_i = (f_{i1},\ldots,f_{ik})^\top \mbox{ with } \boldsymbol{f}_i  \overset{ind}{\sim} N(0,I_k), \\
& \rho_{ij} = \frac{(1-z_{ij}) \exp{(\beta_{0j}+\boldsymbol{f}_i^\top \bbeta_j)}}{\sum_{l=1}^p (1-z_{il}) \exp{(\beta_{0l}+\boldsymbol{f}_i^\top \bbeta_l)}} \\
& \bbeta_j = (\beta_{j1},\ldots,\beta_{jk}) \overset{ind}{\sim} N(0,\Sigma_{\bbeta_j}) \mbox{ and } \eta_j \overset{ind}{\sim} Beta(\alpha_1,\alpha_2) \\
& \bx_i | \brho_i,M_i  \overset{ind}{\sim} MN(\brho_i,M_i),
\end{flalign*}
where ``ind'' stands for independently distributed; Bern, N, Beta and MN denote respectively the Bernoulli, multivariate normal, beta and multinomial distributions. The latent variables $z_{ij}$ model the excess of zeros in the count data; i.e. $z_{ij} = 1$ if $x_{ij} = 0$ with probability $\eta_j$ and $z_{ij} = 0$ if $x_{ij} >0$ with probability $1-\eta_j$. The parameter $\eta_j$ is called the probability of zero-inflation. It is worth pointing out that in our modeling framework, the microbial composition $\brho_i$ does not follow a LN distribution. This is because that the multivariate assumption on the transformed data is not satisfied since the product of two normal distributions is not normal. However, we conjecture that the distribution of $\brho_i$ induces by this log-ratio transformation behaves in a similar way as the LN distribution for overcoming over-dispersion in microbiome count data. Here, the latent vectors $\boldsymbol{f}_i$ represent unobserved environmental factors and the parameter vectors $\bbeta_j$ denote the corresponding factor loadings. Note that the terms $\boldsymbol{f}_i^\top \bbeta_j$ capture the correlations across microbial taxa.      

Let us add two more comments. First, this model is a full Bayesian version of the zero-inflated model proposed by \cite{zeng2022zero} in the sense that no prior distributions on the parameters $\bbeta_j$ and $\eta_j$ for $j=1,\ldots,p$ were assumed therein. More precisely, the authors have set these parameters to specific values in their model setting. Therefore, our inference problem becomes more complex than that developed in \cite{zeng2022zero}; the evidence lower bound is now more difficult to derive for instance. Second, the number of factors or rank, $k$, is assumed to be known in our modeling framework. If there is no prior information about $k$, which is most often the case, standard approach is used to select $k$. Popular methods include cross-validation and information criteria such as the Bayesian information criteria (BIC). For instance, \cite{zeng2022zero} have used the BIC type criterion to select the number of factors $k$. We claim that this approach can be directly extended to our modeling framework without any additional works, however we decide not include this in the current paper. Furthermore, we choose $k$ to be relatively small compared to $p$ i.e. $k \ll p$. This latter point yields a more parsimonious explanation of the dependences between microbial taxa.           

Finally, we call our proposed model ZIPPCA-LPNM. Note its resemblance in nomenclature with ZIPPCA-LNM proposed by \cite{zeng2022zero}. This name is inspired from the log-ratio transformation: since the multivariate normality assumption on the transformed data does not hold, the resulting distribution induced by the inverse transformation is not a LN distribution as in \cite{zeng2022zero}; so we refer to this distribution as the logarithm of the product of two normal distributions (LPN). Again as in \cite{zeng2022zero}, the underlying compositions are given by
\[
\rho_{0ij} = \frac{ \exp{(\beta_{0j}+\boldsymbol{f}_i^\top \bbeta_j)}}{\sum_{j=1}^p \exp{(\beta_{0j}+\boldsymbol{f}_i^\top \bbeta_j)}}.
\]

 \subsection{Variational lower bound}

Let $\bbeta_0 = (\beta_{01},\ldots,\beta_{0p})^\top$, $\bB = (\bbeta_1,\ldots,\bbeta_p)^\top$, $\boldsymbol{\eta} = (\eta_1,\ldots,\eta_p)^\top$ and $\bTheta = \{ \bbeta_0, \bB, \boldsymbol{\eta} \}$. We denote by $l(\bTheta)$ the log-likelihood function for ZIPPCA-LPNM. We let $\hat{\bTheta} = \argmax_{\bTheta} l(\bTheta)$. Note that computing $l(\bTheta)$ requires to sum or integrate over all the latent variables, which can be quite expensive (i.e in the millions or billions of computations) from a computational point of view for complex models and large-scale applications. Exact inference (i.e. inference based on $l(\bTheta)$) is therefore typically intractable in these models, and approximations are needed. In this paper, we use a variational approximation (VA) approach \citep{ormerod2010explaining,blei2017variational} to tackle the inference problem.

The full joint log-distribution for the model is     
\begin{align}
\log p(\bx,\bz,\bB,\boldsymbol{\eta},\boldsymbol{f}) = & \sum_{i=1}^n \log p(\bx_i | \brho_i,M_i) + \sum_{i=1}^n \sum_{j=1}^p \log p(z_{ij} | \eta_j) + \sum_{j=1}^p \log p(\bbeta_j) \nonumber \\
 & + \sum_{j=1}^p \log p(\eta_j) + \sum_{i=1}^n \log p(\boldsymbol{f}_i), 
\end{align}
where
\begin{align*}
\log p(\bx_i | \brho_i,M_i) = & \, \sum_{j=1}^p x_{ij} \log \rho_{ij} + \mbox{Const.} \\
\log p(z_{ij} | \eta_j) = & \, z_{ij} \log \eta_j + (1-z_{ij}) \log (1-\eta_j) \\
\log p (\bbeta_j) = & \, - \frac12 \log | \Sigma_{\bbeta_j} | - \frac12 \bbeta_j^\top \Sigma_{\bbeta_j}^{-1} \bbeta_j + \mbox{Const.} \\
\log p(\eta_j) = & \, (\alpha_1-1) \log \eta_j +(\alpha_2 -1) \log(1-\eta_j) - \log B(\alpha_1,\alpha_2), \\
\log p(\boldsymbol{f}_i) = & \, - \frac12 \boldsymbol{f}_i^\top \boldsymbol{f}_i + \mbox{Const.} 
\end{align*}  
Let $q(\bB,\boldsymbol{\eta},\bz,\boldsymbol{f})$ be an arbitrary density function of the model parameters $(\bB,\boldsymbol{\eta})$ and the latent variables $(\boldsymbol{f},\bz)$. The Kullback-Leibler (KL) divergence between $q(\bB,\boldsymbol{\eta},\bz,\boldsymbol{f})$ and the posterior $p(\bB,\boldsymbol{\eta},\bz,\boldsymbol{f} | \bx)$ satisfies
\begin{align*}
\mbox{KL} \{ q(\bB,\boldsymbol{\eta},\bz,\boldsymbol{f}), p(\bB,\boldsymbol{\eta},\bz,\boldsymbol{f} | \bx) \} = \log p(\bx) - \EE_q [ \log p(\bx,\bz,\bB,\boldsymbol{\eta},\boldsymbol{f}) - \log q(\bB,\boldsymbol{\eta},\bz,\boldsymbol{f}) ],
\end{align*} 
and hence,
\begin{align*}
l(\bTheta) \geq \EE_q [ \log p(\bx,\bz,\bB,\boldsymbol{\eta},\boldsymbol{f}) - \log q(\bB,\boldsymbol{\eta},\bz,\boldsymbol{f}) ],
\end{align*}
where the right-hand side of the inequality is known as ``evidence lower bound'' (ELBO) and which we denote by $\mbox{ELBO}(q)$ in the present paper. The ELBO can also be derived from $\log p(\bx)$ using Jensen's inequality. It is worth pointing out that minimizing the KL divergence between $q$ and $p$ is equivalent to maximizing the ELBO, where $q$ and $p$ replace $q(\bB,\boldsymbol{\eta},\bz,\boldsymbol{f})$ and $p(\bB,\boldsymbol{\eta},\bz,\boldsymbol{f} | \bx)$.

As an approximating variational family, we consider the mean field type family by assuming that
\begin{align*}
q(\bbeta_j,\boldsymbol{\eta},\bz_i,\boldsymbol{f}_i) = q(\bbeta_j) \cdot q(\boldsymbol{f}_i) \cdot \prod_{j=1}^p q(z_{ij}) \cdot \prod_{j=1}^p q(\eta_j)
\end{align*}

where $q(\bbeta_j) \sim N(\br_j,\bLambda_j)$ with $\br_j = (r_{j1},\ldots,r_{jk})^\top$ and $\bLambda_j = \mbox{diag}(\lambda_{j1}^2,\ldots,\lambda_{jk}^2)$, $q(\eta_j) \sim Beta(\gamma_{j1},\gamma_{j2})$, $q(z_{ij}) \sim Bern(\pi_{ij})$, and $q(\boldsymbol{f}_i) \sim N(\bM_i,\bSigma_i)$ with $\bM_i = (m_{i1},\ldots,m_{ik})^\top$ and $\bSigma_i = \mbox{diag}(\sigma_{i1}^2,\ldots,\sigma_{ik}^2)$. Note that in this setting, the parameters $\pi_{ij},\br_j,\bLambda_j,\bM_i,\bSigma_i,\gamma_{j1}$ and $\gamma_{j2}$ 
are called the variational parameters and we denote by $\bDelta$ the set of all variational parameters of our modeling framework. The variational lower bound of the complete data log-likelihood can be written as   

\begin{align*}
\mbox{ELBO}(q) = & - \frac12 \sum_{j=1}^p \left\{ \mbox{trace} \left( \Sigma_{\bbeta_j}^{-1}  (\br_j \br_j^\top + \bLambda_j) \right) - \log | \bLambda_j | \right \} \\
& - \frac12 \sum_{i=1}^n \left\{ \mbox{trace} (\bM_i \bM_i^\top + \bSigma_i) - \log |\bSigma_i | \right\} + \sum_{j=1}^p \log B(\gamma_{j1},\gamma_{j2}) \\
& - \sum_{i=1}^n M_i \log \left\{ \sum_{j=1}^p (1-\pi_{ij}) \exp{(\beta_{0j}+\bL_{ij})}\right \}  - p \log B(\alpha_1,\alpha_2) \\
+ & \sum_{i=1}^n \sum_{j=1}^p \Big\{ \pi_{ij} (\psi(\gamma_{j1})-\psi(\gamma_{j1}+\gamma_{j2})-\log \pi_{ij}) + \\
& (1-\pi_{ij}) (\psi(\gamma_{j2})-\psi(\gamma_{j1}+\gamma_{j2})-\log (1-\pi_{ij})) + x_{ij}(\beta_{0j}+\bM_i^\top \br_j) \Big\} \\
& + \sum_{j=1}^p \Big[ (\alpha_1 - \gamma_{j1}) \{ \psi(\gamma_{j1})-\psi(\gamma_{j1}+\gamma_{j2}) \} + \\
& (\alpha_2 - \gamma_{j2}) \{ \psi(\gamma_{j2})-\psi(\gamma_{j1}+\gamma_{j2}) \} \Big], 
\end{align*}

where $B(\cdot,\cdot)$ is the beta function, $\psi$ is the digamma function, and 

\begin{align*}
\bL_{ij} := & - \frac12 \log \left| \bI_k - \bSigma_i \bLambda_j \right| + \frac12 (2\bM_i + \bSigma_i \br_j)^\top (\bI_k - \bSigma_i \bLambda_j)^{-1} \br_j \\
&+ \frac12 \bM_i^\top \bLambda_j (\bI_k - \bSigma_i \bLambda_j)^{-1} \bM_i.
\end{align*}

\subsection{Parameter estimation and implementation}

For the estimation purpose, we first maximize $\mbox{ELBO}(q)$ over the variational parameters $\bDelta$ to find the VA estimate and to get the variational log-likelihood. Then, we maximize this variational log-likelihood over parameter $\bbeta_0$ to obtain a VA solution for maximum likelihood estimation. Since $\mbox{ELBO}(q)$ does not involve the model parameters $\eta_j$ and $\bbeta_j$, we use the posterior mean as estimates of $\eta_j$ and $\bbeta_j$ i.e.
\begin{align*}
\hat{\eta}_j = \frac{\hat{\gamma}_{j1}}{\hat{\gamma}_{j1}+\hat{\gamma}_{j2}} \mbox{ and } \hat{\bbeta}_j = \hat{\br}_j.
\end{align*}  

The estimate of the compositions $\rho_{ij}$ is then given by 
\begin{align*}
\hat{\rho}_{ij} =  \frac{\exp{(\hat{\beta}_{0j}+\hat{\bM}_i^\top \hat{\br}_j)}}{\sum_{j=1}^p \exp{(\hat{\beta}_{0j}+\hat{\bM}_i^\top \hat{\br}_j)}}.
\end{align*}
In Appendix \ref{multi:poi}, we establish the equivalence between our proposed model and the ZIPPCA-Poisson model. As a consequence, we use this equivalence to update the variational parameter $\pi_{ij}$ 
\[ \hat{\pi}_{ij} = \begin{cases} 
      \frac{\exp(\psi(\gamma_{j1}))}{\exp(\psi(\gamma_{j1})) + \exp(\psi(\gamma_{j2}))\exp(-\exp(\alpha_{i0}+\beta_{0j}+ \bL_{ij}))} & \text{ if } x_{ij} = 0\\
      0 & \text{ if } x_{ij} > 0,
   \end{cases}
\]
where 
\begin{align*}
\alpha_{i0} =  \log \left\{ \frac{M_i}{\sum_{j=1}^p (1-\pi_{ij}) \exp(\beta_{0j} + \bL_{ij})} \right\}.
\end{align*}
As suggested by \cite{zeng2022zero} in an empirical Bayes approach, we also add a classification step on the basis of $\hat{\pi}_{ij}$ by  setting
\[
\pi_{ij}^{(s)} = I(\hat{\pi}_{ij} \geq \pi_0),
\]
where $\pi_0$ is a prespecified threshold with default value 0.5 in our implementation of the methodology in the \texttt{R} software. For more details on the optimization tricks (in the update of $\pi_{ij}$) used to increase speed and convergence, we refer the reader to  \cite{zeng2022zero}, \cite{baker1994multinomial} and \cite{celeux1992classification}. The whole procedure is summarized in Algorithm \ref{alg:algorithm1} and implemented in \texttt{R} software. More details including the optimization problems as well as corresponding objective and gradient functions are given in Appendix \ref{opt:pr}.

To end this section, we add two more comments. First, the computational time of our estimation procedure is comparable to that of \cite{zeng2022zero} in the sense that the convergence rates of two algorithms are on the one hand dimension-dependent and similar (i.e. lie in the same interval of values) on the other hand. More precisely, if the sample size $n$ and the dimension $p$ are high, the computational time induced by the fitting of the corresponding count data tend to be high. In addition, the computational time of the estimation procedure developed in \cite{zeng2022zero} is most often smaller than the one provided by our modeling framework. Second, Algorithm \ref{alg:algorithm1} is always guaranteed to converge. This is because the sequence values of the variational lower is non-decreasing by construction and bounded above by the marginal likelihood or evidence $l(\bTheta)$. The variational lower bound (or more precisely its sequence values) is therefore guaranteed to converge to some limiting value. Note that the different optimization problems that are involved in the estimation procedure are not concave, so for each (variational) parameter there will generally be multiple local maxima as candidate to estimate the corresponding (variational) parameter. In addition, we have no guarantee that our proposed algorithm will find the largest of these local maxima even if it converges within a reasonable time for the experiments that we have considered in the present paper.

\begin{algorithm}
\caption{\enskip Classification Variational Approximation}\label{alg:algorithm1}
\begin{algorithmic}
\State{Initialize \{$\beta_{0j}^{(0)},\br_j^{(0)},\bLambda_j^{(0)},\gamma_{j1}^{(0)},\gamma_{j2}^{(0)},\bM_i^{(0)},\bSigma_i^{(0)},\pi_{ij}^{(0)}$\}, and $s=0$};
  \While{ELBO not converged}
        \State {$s = s+1$};
        \State {given \{$\beta_{0j}^{(s-1)},\gamma_{j1}^{(s-1)},\gamma_{j2}^{(s-1)},\br_j^{(s-1)},\bLambda_j^{(s-1},\bM_i^{(s-1)},\bSigma_i^{(s-1)}$\}, update $\pi_{ij}^{(s)} = I(\hat{\pi}_{ij} \geq \pi_0)$};
        \State {given \{$\beta_{0j}^{(s-1)},\pi_{ij}^{(s)}$\}, update $\gamma_{j1}^{(s)},\gamma_{j2}^{(s)},\br_j^{(s)},\bLambda_j^{(s},\bM_i^{(s)},\bSigma_i^{(s)}$};
        \State {given \{$\br_j^{(s)},\bLambda_j^{(s)},\bM_i^{(s)},\bSigma_i^{(s)},\pi_{ij}^{(s)}$ \}, update $\beta_{0j}^{(s)}$};
    \EndWhile
\end{algorithmic}
\end{algorithm}


\section{Simulation study}\label{sec3}

\subsection{Simulation set-up}
\label{sim:set-up}
We conduct a simulation study to compare the performance of our proposed model to that of the model introduced by \cite{zeng2022zero} in terms of prediction. Specifically, we assess performance by looking at the square root of mean squared error (RMSE) of the parameters. To this end, two scenarios are considered, where count data are sampled from zero-inflated models. More precisely, the setting for generating the count data in Scenario 1 is similar to that of \cite{zeng2022zero} while the count data in Scenario 2 are generated from the ZIPPCA-LPNM model. To investigate the effects of the number of latent variables (k) in each scenario, we consider five different values of $k$, each with two instances of the combinations $(n,p)$ where $n$ is the sample size and $p$ is the dimension. That is, we consider $k=2,5,8,11,14$, $(n,p)=(50,100)$, and $(n,p)=(100,50)$. The intercepts $\beta_{0j}$ are set to $2$, and the sequencing depths $M_i$ are generated from $\mathcal{U}(800,1000)$. Furthermore, the elements of loading vectors $\bbeta_j$ in Scenario 1 are sampled from $\mathcal{U}(-1,1)$ while those in Scenario 2 are generated from $\mathcal{N}(0,0.1\bI)$. For Scenario 1, we set the probabilities of zero-inflation $\eta_j$ at $0.25$ (level of zero-inflation is about $50\%$), and $\eta_j$ are sampled from $\text{Beta}(2,3)$ for Scenario 2. Note that in each scenario, 1000 datasets (count data) are generated from the zero-inflated models. Details on the simulation setup are summarized in Table \ref{table:sim-setup}.


\setlength{\tabcolsep}{1pt}
\renewcommand{\arraystretch}{1.5}
\begin{table}[h]
\caption{Specification of parameters in the zero-inflated models investigated in Section \ref{sim:set-up}. Scenario 1 corresponds to the setting of \cite{zeng2022zero} while Scenario 2 concerns the modeling framework of the present paper.}
\label{table:sim-setup} 
\centering
\begin{tabular}{@{\extracolsep{15pt}} cccccc @{}}
\hline
\hline 
$k$ & $M_i$ & $\beta_{0j}$ & $\bbeta_j$ & $\eta_j$ & $\boldsymbol{f}_i$ \\ 
  \hline
  \hline
  & \multicolumn{5}{c}{Scenario 1}\\
   \hline
 2 & $\mathcal{U}(800,1000)$ & 2 & $\mathcal{U}(-1,1)$ & 0.25 & $\mathcal{N}(0,\bI)$ \\
 5 & $\mathcal{U}(800,1000)$ & 2 & $\mathcal{U}(-1,1)$ & 0.25 & $\mathcal{N}(0,\bI)$ \\ 
 8 & $\mathcal{U}(800,1000)$ & 2 & $\mathcal{U}(-1,1)$ & 0.25 & $\mathcal{N}(0,\bI)$ \\
 11 & $\mathcal{U}(800,1000)$ & 2 & $\mathcal{U}(-1,1)$ & 0.25 & $\mathcal{N}(0,\bI)$ \\
 14 & $\mathcal{U}(800,1000)$ & 2 & $\mathcal{U}(-1,1)$ & 0.25 & $\mathcal{N}(0,\bI)$ \\
\hline
   & \multicolumn{5}{c}{Scenario 2} \\
   \hline
   2 & $\mathcal{U}(800,1000)$ & 2 & $\mathcal{N}(0,0.1\bI)$ & $\text{Beta}(2,3)$ & $\mathcal{N}(0,\bI)$ \\ 
   5 & $\mathcal{U}(800,1000)$ & 2 & $\mathcal{N}(0,0.1\bI)$ & $\text{Beta}(2,3)$ & $\mathcal{N}(0,\bI)$ \\
   8 & $\mathcal{U}(800,1000)$ & 2 & $\mathcal{N}(0,0.1\bI)$ & $\text{Beta}(2,3)$ & $\mathcal{N}(0,\bI)$ \\
   11 & $\mathcal{U}(800,1000)$ & 2 & $\mathcal{N}(0,0.1\bI)$ & $\text{Beta}(2,3)$ & $\mathcal{N}(0,\bI)$ \\
   14 & $\mathcal{U}(800,1000)$ & 2 & $\mathcal{N}(0,0.1\bI)$ & $\text{Beta}(2,3)$ & $\mathcal{N}(0,\bI)$ \\
   
\hline
 \end{tabular}
\end{table}

\subsection{Simulation results}

Results are presented in Tables \ref{table:rmse1} and \ref{table:rmse2}. Overall, the simulation results show the improvement of our proposed model (ZIPPCA-LPNM) over the ZIPPCA-LNM model (introduced by  \cite{zeng2022zero}) in the context of the RMSEs. For both scenarios, the RMSEs of the estimators of the factors $\boldsymbol{f}_i$ and the coefficients $\bbeta_j$ under the ZIPPCA-LPNM method are smaller than that obtained under the ZIPPCA-LNM method. The RMSEs of the estimators of the probability of zero-inflation $\eta_j$ under the ZIPPCA-LNM method are slightly smaller than the ones obtained under our proposed method when $k=5,8,11,14$ in the case of Scenario 1 while the inverse occurs when $k=2$. Moreover, in the context of Scenario 2 the RMSEs of the estimators of the zero-inflations are smaller under our proposed method for all values of the number of factors $k$. For the intercept $\beta_{0j}$, the RMSEs are comparable under the two methods for all two scenarios with smaller RMSEs obtained under the ZIPPCA-LNM method.              

The RMSEs tend to increase as $k$ increases under the two methods for all parameters and all two scenarios except that the RMSEs for the factors $\boldsymbol{f}_i$ diminish when $k$ increases for Scenario 1. The increase in the RMSEs can be explained by the complexity induced of adding more parameters in the estimation procedure when $k$ increases. It is worth pointing out that for the factors $\boldsymbol{f}_i$ and the corresponding coefficients $\bbeta_j$ the RMSEs do not vary much when $k$ increases under our proposed method for all two scenarios, they are equal when rounded to the hundredth place. Overall, our model demonstrates superior performance, outperforming even under simulations based on the zero-inflated model introduced in \cite{zeng2022zero}.



\setlength{\tabcolsep}{5pt}
\renewcommand{\arraystretch}{1.5}
\begin{table}[h]
\caption{RMSEs of estimates of the parameters when $(n,p)=(50,100)$. Results are based on 1000 replications (count datasets) from the zero-inflated model introduced in \cite{zeng2022zero} (Scenario 1) and our model (Scenario 2).}
\label{table:rmse1} 
\centering
\begin{tabular}{@{\extracolsep{8pt}}  l | cccc | cccc @{}}
\hline
\hline 
 &    \multicolumn{4}{c}{ZIPPCA-LNM} & \multicolumn{4}{c}{ZIPPCA-LPNM} \\ 
  \cline{1-9} 
   $k$  & $\hat{\bbeta}_0$ & $\hat{\bEta}$ & $\hat{\bB}$ & $\hat{\bF}$   & $\hat{\bbeta}_0$ & $\hat{\bEta}$ & $\hat{\bB}$ & $\hat{\bF}$   \\ 
  \hline
  \hline
  & \multicolumn{8}{c}{Scenario 1}\\
\hline
   2 & 1.0015 & 0.0685 & 0.6003 & 3.4153 & 1.0224 & 0.0671 & 0.5772 & 0.9979 \\ 
   5  & 1.0042& 0.0925 & 0.6154 & 3.0146 & 1.0550 & 0.1149 & 0.5767 & 0.9996 \\
   8  & 1.0093 & 0.1248 & 0.6287 & 2.5643 & 1.0865 & 0.1545 & 0.5774 & 1.0012 \\
   11  & 1.0105 & 0.1556 & 0.6575 & 2.1840 & 1.1157 & 0.2058 & 0.5775 & 1.0005 \\
   14  & 1.0088 & 0.1784 & 0.6885 & 1.9821 & 1.1460 & 0.2538 & 0.5772 & 0.9995 \\
    
    \hline
   & \multicolumn{8}{c}{Scenario 2} \\
   \hline
   2  & 1.0026 & 0.0690 & 0.4009 & 1.6599 & 1.0090 & 0.0653 & 0.3165 & 0.9989 \\ 
   5  & 1.0069 & 0.0746 & 0.3826 & 1.8243 & 1.0210 & 0.0723 & 0.3159 & 0.9978 \\
   8  & 1.0059 & 0.0802 & 0.3778 & 1.8455 & 1.0342 & 0.0699 & 0.3161 & 0.9987 \\
   11  & 1.0053 & 0.0878 & 0.3804 & 1.7902 & 1.0474 & 0.0745 & 0.3161 & 1.0004 \\
   14  & 1.0242 & 0.1089 & 0.3867 & 1.7432 & 1.0605 & 0.0806 & 0.3159 & 1.0007 \\ 
   
    \hline
 \end{tabular}
\end{table}

\setlength{\tabcolsep}{5pt}
\renewcommand{\arraystretch}{1.5}
\begin{table}[h]
\caption{RMSEs of estimates of the parameters when $(n,p)=(100,50)$. Results are based on 1000 replications (count datasets) from the zero-inflated model introduced in \cite{zeng2022zero} (Scenario 1) and our model (Scenario 2).}
\label{table:rmse2} 
\centering
\begin{tabular}{@{\extracolsep{8pt}}  l | cccc | cccc @{}}
\hline
\hline 
 &    \multicolumn{4}{c}{ZIPPCA-LNM} & \multicolumn{4}{c}{ZIPPCA-LPNM} \\ 
  \cline{1-9} 
   $k$  & $\hat{\bbeta}_0$ & $\hat{\bEta}$ & $\hat{\bB}$ & $\hat{\bF}$   & $\hat{\bbeta}_0$ & $\hat{\bEta}$ & $\hat{\bB}$ & $\hat{\bF}$   \\ 
  \hline
  \hline
  & \multicolumn{8}{c}{Scenario 1}\\
\hline
   2 & 1.0007 & 0.0447 & 0.6226 & 2.6851 & 1.0167 & 0.0486 & 0.5774 & 0.9983 \\ 
   5  & 1.0013 & 0.0541 & 0.6350 & 2.3764 & 1.0406 & 0.0651 & 0.5767 & 0.9993 \\
   8  & 1.0014 & 0.0728 & 0.6452 & 2.1324 & 1.0601 & 0.0998 & 0.5774 & 0.9976 \\
   11  & 1.0015 & 0.0950 & 0.6580 & 1.9809 & 1.0748 & 0.1445 & 0.5770 & 0.9993 \\
   14  & 1.0873 & 0.1098 & 0.6672 & 1.8581 & 1.0873 & 0.1829 & 0.5774 & 1.0001 \\
    
    \hline
   & \multicolumn{8}{c}{Scenario 2} \\
   \hline
   2  & 1.0012 & 0.0449 & 0.4279 & 1.5318 & 1.0057 & 0.0441 & 0.3168 & 0.9999 \\ 
   5  & 1.0017 & 0.0460 & 0.3856 & 1.7490 & 1.0138 & 0.0453 & 0.3173 & 0.9998 \\
   8  & 1.0017 & 0.0460 & 0.3754 & 1.7894 & 1.0212 & 0.0452 & 0.3160 & 0.9986 \\
   11  & 1.0023 & 0.0493 & 0.3759 & 1.7400 & 1.0295 & 0.0485 & 0.3161 & 0.9988 \\
   14  & 1.0035 & 0.0496 & 0.3795 & 1.6719 & 1.0363 & 0.0492 & 0.3162 & 0.9995 \\ 
   
    \hline
 \end{tabular}
\end{table}

\section{Application to real datasets}\label{sec4}

In this section, we present results of applying our method to data from the ``Pediatric-onset Multiple Sclerosis'' study. In this study, stool samples were collected along with patient diet data (Block Kids Food Screener) from the Canadian Pediatric Demyelinating Disease Network. From a total of 37 patients, there were 17 with pediatric-onset multiple sclerosis (MS symptom onset $< 18$ years of age) and 20 controls who are not known to have MS. The multidimensional scaling (MDS) with the Bray-Curtis dissimilarity metric and the method developed in \cite{zeng2022zero} are also performed on this data. We analyze the data at the genus level. After removing samples or features whose total abundances are zero, 296 genera and 37 samples are retained. To complement this section, another metagenomic dataset is considered in order to study the efficiency of our proposed method and its two competitors on real-world data sets.  This dataset comes from the study of the microbiome with pediatric Crohn's disease (CD) cohort performed in \cite{gevers2014treatment}. In addition, we analyze this dataset at the family level with 284 samples and 147 taxa. For both datasets, notice that the number of factors $k$ is set to five (5). Hereafter, we refer respectively to these datasets as $D_1$ and $D_2$. To assess the discriminating power of our methodology between the two groups in the disease diagnosis variable, we conduct for each dataset a logistic regression using the five factors obtained from the model fitting as independent variables and then we compute the corresponding area under curve (AUC) values. Furthermore, we have examined the genera/families that are present in each of the five factors and their corresponding phyla.
  
\begin{figure}[!ht]
\begin{center}
\renewcommand{\arraystretch}{0}
\includegraphics[width=1\textwidth]{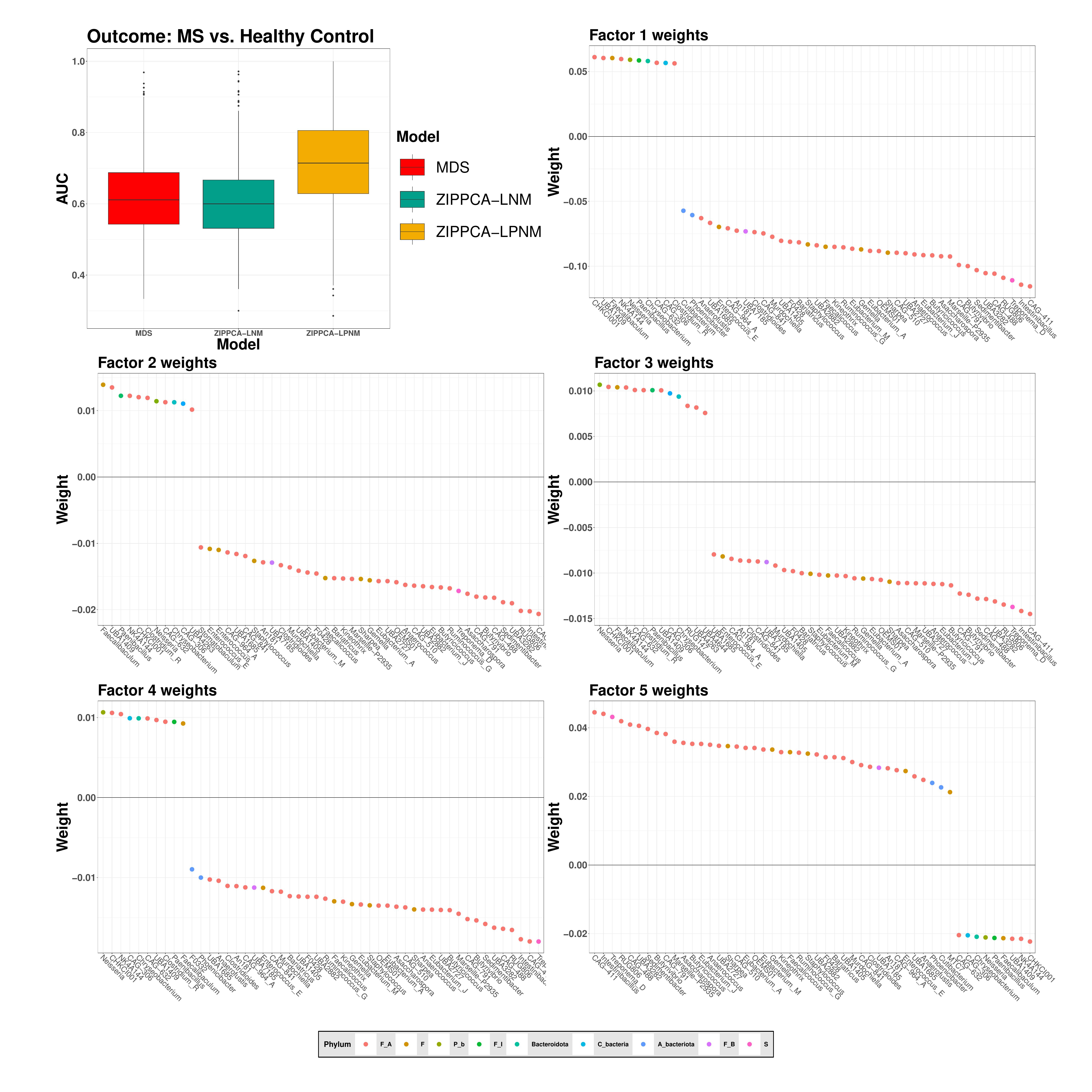}
\caption{Dataset from the Pediatric Multiple Sclerosis study. AUC histogram with logistic regression and genera weights. The logistic regression has disease diagnosis as the response variable and five factors as independent variables. The genera weights indicate the presence of different genera in the five factors. F: Firmicutes; F\_A: Firmicutes\_A; F\_C: Firmicutes\_C; F\_I: Firmicutes\_I; P\_b: Proteobacteria; C\_bacteria: Cyanobacteria;  A\_bacteriota: Actinobacteriota; S: Spirochaetota.}
\label{app:fig1}
\end{center}
\end{figure}

\begin{figure}[!ht]
\begin{center}
\renewcommand{\arraystretch}{0}
\includegraphics[width=1\textwidth]{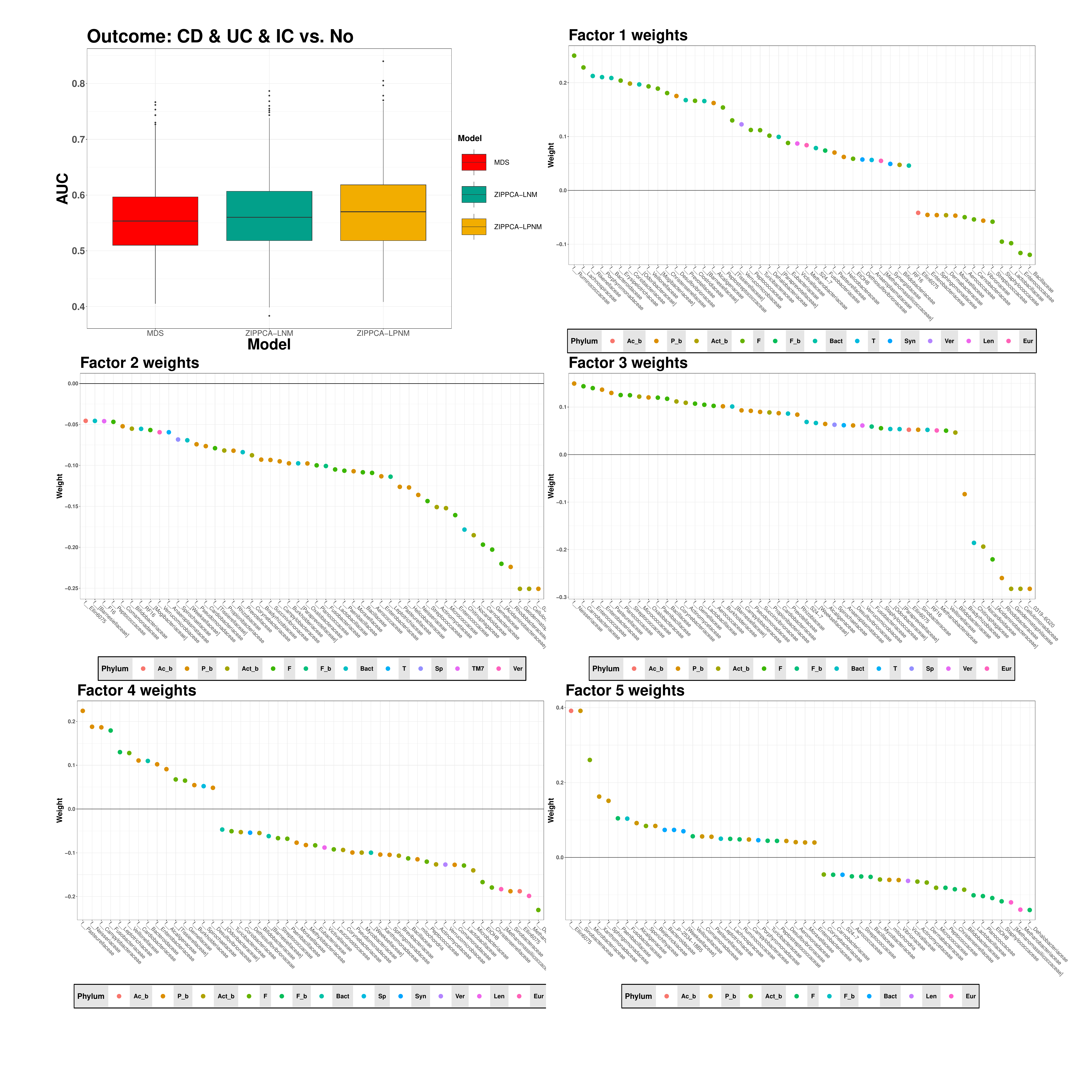}
\caption{Dataset from the study of the microbiome with pediatric CD cohort performed in \cite{gevers2014treatment}. AUC histogram with logistic regression and families weights. The logistic regression has diagnosis (coded as 1 for Crohn's disease (CD), Ulcerative colitis (UC), and Interstitial cystitis (IC), and 0 for `No') as the response variable, along with five factors as independent variables. The families weights indicate the presence of different families in the five factors. Ac\_b: Acidobacteria; P\_b: Proteobacteria; Act\_b: Actinobacteria; F: Firmicutes; Fu\_b: Fusobacteria; Bact: Bacteroidetes; T: Teneicutes; Syn: Synergistetes; Ver: Verrucomicrobia; Len: Lentisphaerae; Eur: Euryarchaeota; Sp: Spirochaetes.}
\label{app:fig2}
\end{center}
\end{figure}

Results are reported in Figures \ref{app:fig1} and \ref{app:fig2}. For each dataset, our method consistently achieves higher AUC values, indicating its superior performance in discriminating between the two groups. For the dataset $D_1$, the genera that are present in the factors have small weights i.e. the weights range between -0.18 and 0.08 approximately. More specifically, the higher weights are obtained with the first factor. In addition, all the selected genera are grouped into the following nine (9) phyla: Actinobacteria, Bacteroidota, Cyanobacteria, Firmicutes, Firmicutes\_A, Firmicutes\_B, Firmicutes\_I, Proteobacteria, and Spirochaetota. The genera weights plots indicate that there is no single phylum that dominates (higher weights) in every factor. This observation suggests that the factors are biologically meaningful to a lesser extent, likely due to the small weights associated with them. Regarding the dataset D2, the families that are present in the factors have relatively high weights, ranging between -0.3 and 0.4 approximately. Note that the first and fifth factors contain families with the highest weights. In addition, the selected families in Figure \ref{app:fig2} fall into the following thirteen (13) phyla: Acidobacteria, Actinobacteria, Bacteroidetes, Euryarchaeota, Firmicutes, Fusobacteria, Lentisphaerae, Proteobacteria, Spirochaetes, Synergistetes, Tenericutes, TM7, and Verrucomicrobia. Across the factors, no single phylum consistently exhibits higher weights, as revealed by the families weights plots. This finding implies that the factors hold biological significance.

\section{Discussion and conclusion}\label{sec:discussion}

In this paper, we develop a fully Bayesian factor analysis technique called Zero-inflated PCA Log-Product-Normal Multinomial (ZIPPCA-LPNM) model that is appropriate for microbiome count data analysis. The developed model enables latent factor analysis. In our setting, we fix the number of (latent) factors to be included in the estimation procedure. In addition, we develop an efficient algorithm based on variational approximation and classification (to a lesser extent) to fit the model. Due to its fast runtimes on large-scale datasets, we use the mean field type family to approximate the marginal likelihood. The computational cost of our methodology is proportional to the size of the count data (i.e. grows as the dimension $p$ increases) and comparable to that introduced in \cite{zeng2022zero}. However, we currently face challenges in managing ultra high-dimensional settings. One way to address this issue may involve leveraging or developing feature screening techniques tailored to the challenges posed by ultra high-dimensional settings, particularly in the context of microbiome count data analysis. The simulation study shows the superior performance of our proposed model over the zero-inflated model introduced in \cite{zeng2022zero} for prediction based on the latent factor analysis and the model parameters. These simulation results also demonstrate a decline in overall model performance as the number of latent factors increases. Finally, we apply our methodology, along with two existing methods, including multidimensional scaling and the zero-inflated model proposed in \cite{zeng2022zero}, to two real datasets. The outcomes reveal that our methodology excels in discriminating between models, surpassing its two competitors based on the area under the curve (AUC) metric. Additionally, the weights of the genera/families present in the latent factors, as provided by our methodology, suggest that these factors carry biological significance. One perspective for this research could involve developing data-driven approaches, such as cross-validation and Bayesian information criteria (BIC), to select the number of latent factors for models within our framework. 

A major drawback of the approach introduced in \cite{zeng2022zero} is that it overestimates the uncertainty of model parameters in the sense that the coverage rates of the model parameters are so much higher than expected, even for models with a small number of factors. Providing an accurate estimate of these uncertainties is essential for making reliable predictions and informed decisions based on the model's outputs. The use of the mean field type family to approximate the posterior distribution enhances computational efficiency and allows for more scalable and efficient Bayesian inference. However, a well known major failing is that the mean field variational Bayes (MFVB) approximation typically underestimates marginal variances \citep{mackay2003information,wang2005inadequacy}. The latter point can be explained by the fact that covariance contributions between the parameters are ignored as the consequence of the assumed independence. By using the sensitivity of MFVB posterior expectations, \cite{giordano2018covariances} developed a methodology that augments MFVB to deliver improved covariance estimates for the model parameters. These sensitivity-based covariance estimates are referred to as linear response variational Bayes (LRVB) covariances; this name and method originated in the statistical mechanics literature \citep{opper2001advanced}. This technique can be used to provide an improved estimator for the exact marginal covariance of the model parameters in our modeling framework. Additionally, it is worth noting that the linear response approach cannot be applied to the setting of the zero-inflated model developed in \cite{zeng2022zero}.

\section*{Author contributions}

\begin{itemize}
\item I. Ba: Conducted the literature review, performed statistical analyses including zero-inflated modeling, parameter estimation, and methodology implementation. Additionally, Ismaila designed and executed a simulation study and applied the methodology to real-world datasets. He then interpreted the results and drew key conclusions. 
\item K. McGregor, M. Turgeon, and S. Muthukumarana: Provided invaluable guidance and supervision throughout the project. 
\item The other authors: Provided the Multiple Sclerosis (MS) dataset essential for the analysis.
\end{itemize}

\section*{Acknowledgments}
The research of I. Ba is supported by the Canadian Statistical Sciences Institute (CANSSI) through the CDPF program. K. McGregor (RGPIN-2021-03634), M. Turgeon (RGPIN-2021-04073), and S. Muthukumarana (RGPIN-2018-05008) all gratefully acknowledge funding via a Discovery Grant from the Natural Sciences and Engineering Research Council of Canada (NSERC). H. Armstrong, S. Veniamin, J. Jovel, and R. Miller all acknowledge funding support from the Canada Research Chair (CRC-2021-00172) program. The authors are grateful for the involvement of all participants, particularly children and youth with multiple sclerosis (MS), and their parents, in the collection of the MS data used in the present paper. The authors are also grateful to the investigators and study teams at each site who participated in the Canadian Paediatric Demyelinating Disease Network study.




\section*{Conflict of interest}

The authors declare no potential conflict of interests.




\newpage

\appendix


\section{Full derivation of variational lower bound for the ZIPPCA-LPNM model}
\label{elbo:zippcalpnm}
\vspace*{12pt}
Recall that $\mbox{ELBO}(q) = \EE_q [ \log p(\bx,\bz,\bB,\boldsymbol{\eta},\boldsymbol{f}) - \log q(\bB,\boldsymbol{\eta},\bz,\boldsymbol{f}) ]$. We split its calculation into two parties.
\begin{enumerate}
\item \underline{Calculation of $\EE_q[\log q(\bB,\boldsymbol{\eta},\bz,\boldsymbol{f})]$}. Note that
\begin{align*}
\log q(\bB,\boldsymbol{\eta},\bz,\boldsymbol{f}) = & \sum_{j=1}^p \log q(\bbeta_j) + \sum_{j=1}^p \log q(\eta_j) + \sum_{i=1}^n \sum_{j=1}^p \log q(z_{ij}) + \\
& \sum_{i=1}^n \log q(\boldsymbol{f}_i),
\end{align*}
where we deduce that
\begin{align*}
\EE_q[\log q(\bB,\boldsymbol{\eta},\bz,\boldsymbol{f})] = & \sum_{j=1}^p \EE_q[\log q(\bbeta_j)] + \sum_{j=1}^p \EE_q[\log q(\eta_j)] +  \\
&\sum_{i=1}^n \sum_{j=1}^p \EE_q[\log q(z_{ij})] + \sum_{i=1}^n \EE_q [\log q(\boldsymbol{f}_i)],
\end{align*} 
with
\begin{align*}
\EE_q[\log q(\bbeta_j)] = & - \frac12 \log | \bLambda_j| + \mbox{Const}, \\
\EE_q[\log q(z_{ij})] = & \pi_{ij} \log \pi_{ij} +(1-\pi_{ij}) \log(1-\pi_{ij}) \\
\EE_q[\log q(\eta_j)] = & (\gamma_{j1}-1) \{ \psi(\gamma_{j1})-\psi(\gamma_{j1}+\gamma_{j2}) \} + (\gamma_{j2}-1) \{ \psi(\gamma_{j2})-\psi(\gamma_{j1}+\gamma_{j2}) \} \\
& - \log B(\gamma_{j1},\gamma_{j2}), \\
\EE_q[\log q(\boldsymbol{f}_i)] = & - \frac12 \log | \bSigma_i| + \mbox{Const},
\end{align*}
where $\psi$ is the digamma function. \\

\item \underline{Calculation of $\EE[\log p(\bx,\bz,\bB,\boldsymbol{\eta},\boldsymbol{f})]$}. We have 
\begin{align*}
\EE_q[\log p(\bx,\bz,\bB,\boldsymbol{\eta},\boldsymbol{f})] = & \sum_{i=1}^n \EE_q[\log p(\bx_i | \brho_i,M_i)] + \sum_{i=1}^n \sum_{j=1}^p \EE_q[\log p(z_{ij} | \eta_j)] \\
& + \sum_{j=1}^p \EE_q[\log p(\bbeta_j)] + \sum_{j=1}^p \EE_q[\log p(\eta_j)] + \sum_{i=1}^n \EE_q[\log p(\boldsymbol{f}_i)],
\end{align*}
where
\begin{align*}
 \EE_q[\log p(z_{ij} | \eta_j)] = & \pi_{ij} \{ \psi(\gamma_{j1})-\psi(\gamma_{j1}+\gamma_{j2}) \} + (1-\pi_{ij}) \{ \psi(\gamma_{j2})-\psi(\gamma_{j1}+\gamma_{j2}) \}, \\
 \EE_q[\log p(\bbeta_j)] = & - \frac12 \sum_{j=1}^p  \mbox{trace} \left( \Sigma_{\bbeta_j}^{-1}  (\br_j \br_j^\top + \bLambda_j) \right) + \mbox{Const}, \\
 \EE_q[\log p(\eta_j)] = & (\alpha_1-1) \{ \psi(\gamma_{j1})-\psi(\gamma_{j1}+\gamma_{j2}) \} + (\alpha_2-1) \{ \psi(\gamma_{j2})-\psi(\gamma_{j1}+\gamma_{j2}) \} \\
 & - \log B(\alpha_1,\alpha_2), \\
 \EE_q[\log p(\boldsymbol{f}_i)] = & - \frac12 \sum_{i=1}^n \mbox{trace} (\bM_i \bM_i^\top + \bSigma_i) + \mbox{Const}, \\
 \EE_q[\log p(\bx_i | \brho_i,M_i)] = & \sum_{j=1}^p x_{ij} (\beta_{0j}+\bM_i^\top \br_j) - M_i \log \left \{ \sum_{j=1}^p \left( (1-\pi_{ij}) \exp(\beta_{0j}) \prod_{l=1}^k L_{ijl} \right) \right \} \\
 = & \sum_{j=1}^p x_{ij} (\beta_{0j}+\bM_i^\top \br_j) - M_i \log \left\{ \sum_{j=1}^p (1-\pi_{ij}) \exp \left( \beta_{0j} + \sum_{l=1}^k \log L_{ijl} \right) \right\} \\
 = &  \sum_{j=1}^p x_{ij} (\beta_{0j}+\bM_i^\top \br_j) - M_i \log  \left\{ \sum_{j=1}^p (1-\pi_{ij}) \exp \left( \beta_{0j} + \bL_{ij} \right) \right\}
 \end{align*}
 
 \begin{align*}
\mbox{with } L_{ijl} = \frac{1}{\sqrt{1 - \sigma_{il}^2 \lambda_{jl}^2}} \exp{ \left(\frac{\sigma_{il}^2 r_{jl}^2+\lambda_{jl}^2 m_{il}^2 + 2 m_{il} r_{jl}}{2(1-\sigma_{il}^2 \lambda_{jl}^2)} \right)} 
 \end{align*}
 is the moment generating function of the product of two normal distributions evaluated at $1$~\citep{seijas2020approximating} and
 \begin{align*}
 \bL_{ij}:=\sum_{l=1}^k \log L_{ijl} = & - \frac12 \log | \bI_k - \bSigma_i \bLambda_j | + \bM_i^\top (\bI_k - \bSigma_i \bLambda_j)^{-1} \br_j + \\
 & \frac12  \br_j^\top \bSigma_i (\bI_k-\bSigma_i \bLambda_j)^{-1} \br_j + \frac12 \bM_i^\top \bLambda_j (\bI_k-\bSigma_i \bLambda_j)^{-1} \bM_i \\
 = & - \frac12 \log | \bI_k - \bSigma_i \bLambda_j | + \frac12 (2 \bM_i + \bSigma_i \br_j)^\top (\bI_k - \bSigma_i \bLambda_j)^{-1} \br_j  \\
 & + \frac12 \bM_i^\top \bLambda_j (\bI_k - \bSigma_i \bLambda_j)^{-1} \bM_i.
 \end{align*}
\end{enumerate}

\section{Optimization problems}%
\label{opt:pr}
\vspace*{12pt}
In this section, we provide the different optimization problems to solve in order to find the estimates of $\bbeta_0$ and variational parameters. For each parameter of interest, we define the optimization problem and provide the corresponding objective and score functions. For updating (or estimating) the parameters, we use the BFGS quasi-Newton method. It is worth pointing out that the elements of $\bLambda_j$ and $\bSigma_i$ are restricted on the interval $(0,1)$ in order to have a well-defined variational lower bound. Consequently, we use the ``L-BFGS-B'' method to update variational parameters $\bsigma_i^2$ and $\blambda_j^2$ for $i=1,\ldots,n$ and $j=1,\ldots,p$. Furthermore, denote elementwise multiplication of two k-vectors $\bu$ and $\bv$ by $\bu \cdot \bv = (u_1 v_1,u_2 v_2,\ldots,u_k v_k)^\top$. For notational convenience, we use this elementwise multiplication operation to define $\bM_i^2:=(m_{i1}^2,\ldots,m_{ik}^2)^\top$, $\br_j^2:=(r_{j1}^2,\ldots,r_{jk}^2)^\top$, and $\br_j \bM_i =  \bM_i \br_j:=(m_{i1}r_{j1},\ldots,m_{ik}r_{jk})^\top$.

\begin{enumerate}
\item Update $\gamma_{j1}^{(s)}$
\begin{align*}
\gamma_{j1}^{(s)} = \argmax_{\gamma_{j1}} f(\gamma_{j1}) \mbox{ with}
\end{align*}

\begin{align*}
f(\gamma_{j1}) = & \sum_{i=1}^n \left\{ \pi_{ij}^{(s)} \left( \psi(\gamma_{j1})-\psi(\gamma_{j1}+\gamma_{j2}^{(s-1)}) \right) - (1-\pi_{ij}^{(s)}) \psi(\gamma_{j1}+\gamma_{j2}^{(s-1)}) \right\} + \\
&  (\alpha_1 - \gamma_{j1}) \left( \psi(\gamma_{j1})-\psi(\gamma_{j1}+\gamma_{j2}^{(s-1)}) \right) -  (\alpha_2 - \gamma_{j2}^{(s-1)}) \psi(\gamma_{j1}+\gamma_{j2}^{(s-1)}) \\
& +  \log B(\gamma_{j1},\gamma_{j2}^{(s-1)}), \\
\frac{\partial f(\gamma_{j1})}{\gamma_{j1}} = & \sum_{i=1}^n  \left\{ \pi_{ij}^{(s)} \left( \psi_1(\gamma_{j1})-\psi_1(\gamma_{j1}+\gamma_{j2}^{(s-1)}) \right) - (1-\pi_{ij}^{(s)}) \psi_1(\gamma_{j1}+\gamma_{j2}^{(s-1)}) \right\} + \\
&  (\alpha_1 - \gamma_{j1}) \left( \psi_1(\gamma_{j1})-\psi_1(\gamma_{j1}+\gamma_{j2}^{(s-1)}) \right) -  (\alpha_2 - \gamma_{j2}^{(s-1)}) \psi_1(\gamma_{j1}+\gamma_{j2}^{(s-1)}).
\end{align*}

\item Update $\gamma_{j2}^{(s)}$
\begin{align*}
\gamma_{j2}^{(s)} = \argmax_{\gamma_{j2}} f(\gamma_{j2}) \mbox{ with}
\end{align*}

\begin{align*}
f(\gamma_{j2}) = & \sum_{i=1}^n \left\{ - \pi_{ij}^{(s)} \psi(\gamma_{j1}^{(s)}+\gamma_{j2})  + (1-\pi_{ij}^{(s)})  \left( \psi(\gamma_{j2})-\psi(\gamma_{j1}^{(s)}+\gamma_{j2}) \right) \right\} + \\
&  (\alpha_2 - \gamma_{j2}) \left( \psi(\gamma_{j2})-\psi(\gamma_{j1}^{(s)}+\gamma_{j2}) \right) -  (\alpha_1 - \gamma_{j1}^{(s)}) \psi(\gamma_{j1}^{(s)}+\gamma_{j2}) \\
& +  \log B(\gamma_{j1}^{(s)},\gamma_{j2}), \\
\frac{\partial f(\gamma_{j2})}{\gamma_{j2}} = & \sum_{i=1}^n \left\{ - \pi_{ij}^{(s)} \psi_1(\gamma_{j1}^{(s)}+\gamma_{j2})  + (1-\pi_{ij}^{(s)})  \left( \psi_1(\gamma_{j2})-\psi_1(\gamma_{j1}^{(s)}+\gamma_{j2}) \right) \right\} + \\
&  (\alpha_2 - \gamma_{j2}) \left( \psi_1(\gamma_{j2})-\psi_1(\gamma_{j1}^{(s)}+\gamma_{j2}) \right) -  (\alpha_1 - \gamma_{j1}^{(s)}) \psi_1(\gamma_{j1}^{(s)}+\gamma_{j2}).
\end{align*}

\item Update $\br_j^{(s)}$

\begin{align*}
\br_j^{(s)} = & \argmax_{\br_j} \Biggl[ - \frac12  \left\{ \mbox{trace} \left( \Sigma_{\bbeta_j}^{-1}  (\br_j \br_j^\top + \bLambda_j^{(s-1)}) \right) - \log | \bLambda_j^{(s-1)} | \right \} + \\
&  \sum_{i=1}^n x_{ij} \br_j^\top \bM_i^{(s-1)} - \sum_{i=1}^n M_i \log \left\{ \sum_{j=1}^p (1-\pi_{ij}^{(s)}) \exp{(\beta_{0j}^{(s-1)}+\bL_{ij}^{(s-1)})}\right \} \Biggr], \\
\end{align*}

\begin{align*}
\frac{\partial \mbox{ELBO}(q)}{\partial \br_j} = & -\sum_{i=1}^n M_i  \left\{  \frac{ (1-\pi_{ij}^{(s)})\bT_{ij}^{(s-1)} \exp{(\beta_{0j}^{(s-1)}+\bL_{ij}^{(s-1)})}}{ \sum_{j=1}^p (1-\pi_{ij}^{(s)}) \exp{(\beta_{0j}^{(s-1)}+\bL_{ij}^{(s-1)})}} \right\}  \\
& - \Sigma_{\bbeta_j}^{-1} \br_j + \sum_{i=1}^n x_{ij} \bM_i^{(s-1)},
\end{align*}

\begin{align*}
\mbox{ where } \bT_{ij}^{(s-1)} = & \frac{\partial \bL_{ij}^{(s-1)}}{\partial \br_j} = \left( \bI_k - \bSigma_i^{(s-1)} \bLambda_j^{(s-1)}\right)^{-1} \bM_i^{(s-1)}  \, + \\
& \bSigma_i^{(s-1)} \left(\bI_k - \bSigma_i^{(s-1)} \bLambda_j^{(s-1)}\right)^{-1} \br_j
\end{align*}

\item Update $\bLambda_j^{(s)}$ $(\blambda_j^2 = (\lambda_{j1}^2,\ldots,\lambda_{jk}^2)^\top)$

\begin{align*}
\blambda_j^{2(s)} = & \argmax_{\blambda_j^2} \Biggl[ - \frac12  \left\{ \mbox{trace} \left( \Sigma_{\bbeta_j}^{-1}   \bLambda_j \right) - \log | \bLambda_j | \right \}  \\
&   - \sum_{i=1}^n M_i \log \left\{ \sum_{j=1}^p (1-\pi_{ij}^{(s)}) \exp{(\beta_{0j}^{(s-1)}+\bL_{ij}^{(s-1)})}\right \} \Biggr], \\
\end{align*}

\begin{align*}
\frac{\partial \mbox{ELBO}(q)}{\partial \blambda_j^2} = & - \frac12 \sum_{i=1}^n M_i  \left\{  \frac{ (1-\pi_{ij}^{(s)})\bW_{ij}^{(s-1)} \exp{(\beta_{0j}^{(s-1)}+\bL_{ij}^{(s-1)})}}{ \sum_{j=1}^p (1-\pi_{ij}^{(s)}) \exp{(\beta_{0j}^{(s-1)}+\bL_{ij}^{(s-1)})}} \right\}  \\
& - \frac12 (\Sigma_{\bbeta_j}^{-1} - \blambda_j^{-2}),
\end{align*}

\begin{align*}
\mbox{ where } \bW_{ij}^{(s-1)} = & 2 \frac{\partial \bL_{ij}^{(s-1)}}{\partial \blambda_j^2} = \left( \bI_k - \bSigma_i^{(s-1)} \bLambda_j^{(s-1)}\right)^{-1} \bsigma_i^{2(s-1)}  \, + \\
&   \left(\bI_k - \bSigma_i^{(s-1)} \bLambda_j\right)^{-2} \bM_i^{2(s-1)} \, + \\
&  \left(\bI_k - \bSigma_i^{(s-1)} \bLambda_j\right)^{-2} \bSigma_i^{(s-1)} \Big( 2 \bM_i^{(s-1)} \br_j^{(s)} + \bSigma_i^{(s-1)} \br_j^{2(s)} \Big).
\end{align*}
 
\item Update $\bM_i^{(s)}$

\begin{align*}
\bM_i^{(s)} = & \argmax_{\bM_i} \Biggl[ - \frac12 \left\{ \mbox{trace} \left(  \bM_i \bM_i^\top + \bSigma_i^{(s-1)} \right) - \log | \bSigma_j^{(s-1)} | \right \} + \\
&  \sum_{j=1}^p x_{ij} \bM_i^\top \br_j^{(s)} -  M_i \log \left\{ \sum_{j=1}^p (1-\pi_{ij}^{(s)}) \exp{(\beta_{0j}^{(s-1)}+\bL_{ij}^{(s-1)})}\right \} \Biggr], \\
\frac{\partial \mbox{ELBO}(q)}{\partial \bM_i} = & - M_i  \left\{  \frac{ \sum_{j=1}^p (1-\pi_{ij}^{(s)})\bU_{ij}^{(s-1)} \exp{(\beta_{0j}^{(s-1)}+\bL_{ij}^{(s-1)})}}{ \sum_{j=1}^p (1-\pi_{ij}^{(s)}) \exp{(\beta_{0j}^{(s-1)}+\bL_{ij}^{(s-1)})}} \right\}  \\
& -  \bM_i + \sum_{j=1}^p x_{ij} \br_j^{(s)},
\end{align*}

\begin{align*}
\mbox{ where } \bU_{ij}^{(s-1)} = & \frac{\partial \bL_{ij}^{(s-1)}}{\partial \bM_i} = \left( \bI_k - \bSigma_i^{(s-1)} \bLambda_j^{(s)}\right)^{-1} \br_j^{(s)}  \, + \\
& \bLambda_j^{(s)} \left(\bI_k - \bSigma_i^{(s-1)} \bLambda_j^{(s)}\right)^{-1} \bM_i.
\end{align*}

\item Update $\bSigma_i^{(s)}$ $(\bsigma_i^2 = (\sigma_{i1}^2,\ldots,\sigma_{ik}^2)^\top)$

\begin{align*}
\bsigma_i^{2(s)} = & \argmax_{\bsigma_i^2} \Biggl[ - \frac12 \left\{ \mbox{trace} \left(  \bSigma_i \right) - \log | \bSigma_i | \right \}  \\
&  -  M_i \log \left\{ \sum_{j=1}^p (1-\pi_{ij}^{(s)}) \exp{(\beta_{0j}^{(s-1)}+\bL_{ij}^{(s-1)})}\right \} \Biggr], \\
\frac{\partial \mbox{ELBO}(q)}{\partial \bsigma_i^2} = & - \frac12 M_i  \left\{  \frac{ \sum_{j=1}^p (1-\pi_{ij}^{(s)})\bS_{ij}^{(s-1)} \exp{(\beta_{0j}^{(s-1)}+\bL_{ij}^{(s-1)})}}{ \sum_{j=1}^p (1-\pi_{ij}^{(s)}) \exp{(\beta_{0j}^{(s-1)}+\bL_{ij}^{(s-1)})}} \right\}  \\
& -  \frac12 (1 - \bsigma_i^{-2}),
\end{align*}

\begin{align*}
\mbox{ where } \bS_{ij}^{(s-1)} = 2 \frac{\partial \bL_{ij}^{(s-1)}}{\partial \bsigma_i^2} = & \left( \bI_k - \bSigma_i \bLambda_j^{(s)}\right)^{-1} \blambda_j^{2(s)}  \, + \\
&    \left(\bI_k - \bSigma_i \bLambda_j^{(s)}\right)^{-2} \bLambda_j^{2(s)} \bM_i^{2(s)}  
\\
&  +    \left(\bI_k - \bSigma_i \bLambda_j^{(s)}\right)^{-2} \Big( \br_j^{2(s)} + 2 \bLambda_j^{(s)} \br_j^{(s)}\bM_i^{(s)} \Big).
\end{align*}

\item Update $\beta_{0j}^{(s)}$

\begin{align*}
\beta_{0j}^{(s)} = \argmax_{\beta_{0j}} f(\beta_{0j}) \mbox{ with}
\end{align*}

\begin{align*}
f(\beta_{0j}) = & \sum_{i=1}^n x_{ij} \beta_{0j} -  \sum_{i=1}^n M_i \log \left\{ \sum_{j=1}^p (1-\pi_{ij}^{(s)}) \exp{(\beta_{0j}+\bL_{ij}^{(s)})}\right \}, \\
\frac{\partial f(\beta_{0j})}{\partial \beta_{0j}} = & \sum_{i=1}^n x_{ij} -  \sum_{i=1}^n M_i  \left\{  \frac{ (1-\pi_{ij}^{(s)})\exp{(\beta_{0j}+\bL_{ij}^{(s)})} }{ \sum_{j=1}^p (1-\pi_{ij}^{(s)}) \exp{(\beta_{0j}+\bL_{ij}^{(s)})}} \right\}.
\end{align*}

\end{enumerate}

\section{ELBO for ZIPPCA-Poisson model}%
\label{multi:poi}
\vspace*{12pt}
The following model is related to the ZIPPCA-LPNM model. It is used for instance to update the variational parameter $\pi_{ij}$. Let 
\begin{flalign*}
&\mbox{latent space } z_{ij} \sim Bern(\eta_j), \boldsymbol{f}_i \sim N(0,\bI_k), \\
&\mbox{parameter space } \log \mu_{ij} = \alpha_{i0} + \beta_{0j} \boldsymbol{f}_i^\top \bbeta_j, \\
& \hspace{2.7cm} \bbeta_j \sim N(0,\Sigma_{\bbeta_j}),  \eta_j \sim Beta(\alpha_1,\alpha_2), \\
& \mbox{observation space } \bx_i | \mu_{ij},z_{ij} = \begin{cases} 
      0 & \mbox{ if } z_{ij} = 1\\
      Pois(\mu_{ij}) & \mbox{ if } z_{ij} = 0.
   \end{cases}
\end{flalign*}

The corresponding variational lower bound writes

\begin{align*}
\mbox{ELBO}(q) = & - \frac12 \sum_{j=1}^p \left\{ \mbox{trace} \left( \Sigma_{\bbeta_j}^{-1}  (\br_j \br_j^\top + \bLambda_j) \right) - \log | \bLambda_j | \right \} \\
& - \frac12 \sum_{i=1}^n \left\{ \mbox{trace} (\bM_i \bM_i^\top + \bSigma_i) - \log |\bSigma_i | \right\} + \\
 & \sum_{i=1}^n \sum_{j=1}^p \Big\{ \pi_{ij} (\psi(\gamma_{j1})-\psi(\gamma_{j1}+\gamma_{j2})-\log \pi_{ij}) + \\
& \hspace{-1cm} (1-\pi_{ij}) (\psi(\gamma_{j2})-\psi(\gamma_{j1}+\gamma_{j2})-\log (1-\pi_{ij})) + x_{ij}(\alpha_{i0} + \beta_{0j}+\bM_i^\top \br_j) \\
 & - \exp(\alpha_{i0} + \beta_{0j} + \bL_{ij}) - \log(x_{ij}!) \Big\} \, + \\
& \hspace{-1.8cm}  \sum_{j=1}^p \Big[ (\alpha_1 - \gamma_{j1}) \{ \psi(\gamma_{j1})-\psi(\gamma_{j1}+\gamma_{j2}) \} + (\alpha_2 - \gamma_{j2}) \{ \psi(\gamma_{j2})-\psi(\gamma_{j1}+\gamma_{j2}) \} \Big].
\end{align*}

Now, taking the first and second derivatives of $\mbox{ELBO}(q,\alpha_{i0})$ with respect to $\alpha_{i0}$, we obtain
\begin{align*}
\frac{\partial \mbox{ELBO}(q,\alpha_{i0})}{\partial \alpha_{i0}} = & \sum_{j=1}^p (1-\pi_{ij}) \{ x_{ij} - \exp(\alpha_{i0} + \beta_{0j} + \bL_{ij}) \}, \\
\frac{\partial^2 \mbox{ELBO}(q,\alpha_{i0})}{\partial \alpha_{i0}^2} = & - \sum_{j=1}^p (1-\pi_{ij}) \exp(\alpha_{i0} + \beta_{0j} + \bL_{ij}).
\end{align*}

Thus, $\mbox{ELBO}(q,\alpha_{i0})$ has a unique maximizer 
\begin{align*}
\hat{\alpha}_{i0} = & \log \left\{ \frac{\sum_{j=1}^p (1-\pi_{ij}) x_{ij}}{\sum_{j=1}^p (1-\pi_{ij}) \exp( \beta_{0j} + \bL_{ij})} \right\} \\
= & \log \left\{ \frac{M_i}{\sum_{j=1}^p (1-\pi_{ij}) \exp( \beta_{0j} + \bL_{ij})} \right\}.
\end{align*} 

Therefore, plugging in $\hat{\alpha}_{i0}$ as a substitute for $\alpha_{i0}$ yields

\begin{align*}
\mbox{ELBO}(q) = & - \frac12 \sum_{j=1}^p \left\{ \mbox{trace} \left( \Sigma_{\bbeta_j}^{-1}  (\br_j \br_j^\top + \bLambda_j) \right) - \log | \bLambda_j | \right \} \\
& - \frac12 \sum_{i=1}^n \left\{ \mbox{trace} (\bM_i \bM_i^\top + \bSigma_i) - \log |\bSigma_i | \right\} - \sum_{i=1}^n M_i \\
 & + \sum_{i=1}^n \sum_{j=1}^p \Big\{ \pi_{ij} (\psi(\gamma_{j1})-\psi(\gamma_{j1}+\gamma_{j2})-\log \pi_{ij}) + \\
& \hspace{-1cm} (1-\pi_{ij}) (\psi(\gamma_{j2})-\psi(\gamma_{j1}+\gamma_{j2})-\log (1-\pi_{ij})) + x_{ij}( \beta_{0j}+\bM_i^\top \br_j) \Big\} + \\
&  \hspace{-2cm} \sum_{j=1}^p \Big[ (\alpha_1 - \gamma_{j1}) \{ \psi(\gamma_{j1})-\psi(\gamma_{j1}+\gamma_{j2}) \} + (\alpha_2 - \gamma_{j2}) \{ \psi(\gamma_{j2})-\psi(\gamma_{j1}+\gamma_{j2}) \} \Big] \\
& \hspace{-2cm} - \sum_{i=1}^n M_i \log  \left\{ \sum_{j=1}^p (1-\pi_{ij}) \exp \left( \beta_{0j} + \bL_{ij} \right) \right\} + \sum_{i=1}^n \sum_{j=1}^p \left[ x_{ij} \log M_i - \log (x_{ij}!) \right].
\end{align*}

This implies an equivalence between the ELBO of ZIPPCA-Poisson and that of ZIPPCA-LPNM. The first and second derivatives of $\text{ELBO}(q,\alpha_{i0})$ with respect to $\pi_{ij}$ write
\begin{align*}
\frac{\partial \text{ELBO}(q,\alpha_{i0})}{\partial \pi_{ij}} = & \psi(\gamma_{j1}) - \psi(\gamma_{j2}) - \log \left( \frac{\pi_{ij}}{1-\pi_{ij}} \right) - x_{ij} (\alpha_{i0} + \beta_{j0} + \bM_i^\top \br_j) + \\
& \exp(\alpha_{i0}+\beta_{0j}+ \bL_{ij}) + \log(x_{ij}!) \\
\frac{\partial^2 \text{ELBO}(q,\alpha_{i0})}{\partial \pi_{ij}^2} = & -\frac{1}{\pi_{ij}} - \frac{1}{1-\pi_{ij}}.
\end{align*}

Thus, $\text{ELBO}(q,\alpha_{i0})$ has a unique maximizer
\[ \hat{\pi}_{ij} = \begin{cases} 
      \frac{\exp(\psi(\gamma_{j1}))}{\exp(\psi(\gamma_{j1})) + \exp(\psi(\gamma_{j2}))\exp(-\exp(\alpha_{i0}+\beta_{0j}+ \bL_{ij}))} & \text{ if } x_{ij} = 0\\
      0 & \text{ if } x_{ij} > 0.
   \end{cases}
\] 

Plugging this into $\text{ELBO}(q,\alpha_{i0})$, one obtains  

\begin{align*}
\mbox{ELBO}(q) = & - \frac12 \sum_{j=1}^p \left\{ \mbox{trace} \left( \Sigma_{\bbeta_j}^{-1}  (\br_j \br_j^\top + \bLambda_j) \right) - \log | \bLambda_j | \right \} \\
& - \frac12 \sum_{i=1}^n \left\{ \mbox{trace} (\bM_i \bM_i^\top + \bSigma_i) - \log |\bSigma_i | \right\} + \\
 & \hspace{-2cm} \sum_{i=1}^n \sum_{j:x_{ij}=0} \Big\{ \log \left[ {\exp(\psi(\gamma_{j1})) + \exp(\psi(\gamma_{j2}))\exp(-\exp(\alpha_{i0}+\beta_{0j}+ \bL_{ij}))} \right] - \\
 & \hspace{-2cm}  \psi(\gamma_{j1}+\gamma_{j2}) \Big\} + \sum_{i=1}^n \sum_{j:x_{ij}>0} \Big\{ \psi(\gamma_{j2}) - \psi(\gamma_{j1}+\gamma_{j2}) + x_{ij} (\alpha_{i0} + \beta_{j0} + \bM_i^\top \br_j) - \\
 &  \hspace{-2cm} \exp(\alpha_{i0}+\beta_{0j}+ \bL_{ij}) - \log(x_{ij}! \Big\} +   \sum_{j=1}^p \Big\{ (\alpha_1 - \gamma_{j1}) \{ \psi(\gamma_{j1})-\psi(\gamma_{j1}+\gamma_{j2}) \} \, + \\
& (\alpha_2 - \gamma_{j2}) \{ \psi(\gamma_{j2})-\psi(\gamma_{j1}+\gamma_{j2}) \} \Big\}.
\end{align*}


\bibliographystyle{imsart-nameyear}
\bibliography{BFA}

\begin{thebibliography}{24}

\bibitem[\protect\citeauthoryear{Aitchison}{1982}]{aitchison1982statistical}
\begin{barticle}[author]
\bauthor{\bsnm{Aitchison},~\bfnm{John}\binits{J.}}
(\byear{1982}).
\btitle{The statistical analysis of compositional data}.
\bjournal{Journal of the Royal Statistical Society: Series B (Methodological)}
\bvolume{44}
\bpages{139--160}.
\end{barticle}
\endbibitem

\bibitem[\protect\citeauthoryear{Baker}{1994}]{baker1994multinomial}
\begin{barticle}[author]
\bauthor{\bsnm{Baker},~\bfnm{Stuart~G}\binits{S.~G.}}
(\byear{1994}).
\btitle{The multinomial-Poisson transformation}.
\bjournal{Journal of the Royal Statistical Society: Series D (The
  Statistician)}
\bvolume{43}
\bpages{495--504}.
\end{barticle}
\endbibitem

\bibitem[\protect\citeauthoryear{Billheimer, Guttorp and
  Fagan}{2001}]{billheimer2001statistical}
\begin{barticle}[author]
\bauthor{\bsnm{Billheimer},~\bfnm{Dean}\binits{D.}},
  \bauthor{\bsnm{Guttorp},~\bfnm{Peter}\binits{P.}} \AND
  \bauthor{\bsnm{Fagan},~\bfnm{William~F}\binits{W.~F.}}
(\byear{2001}).
\btitle{Statistical interpretation of species composition}.
\bjournal{Journal of the American statistical Association}
\bvolume{96}
\bpages{1205--1214}.
\end{barticle}
\endbibitem

\bibitem[\protect\citeauthoryear{Blei, Kucukelbir and
  McAuliffe}{2017}]{blei2017variational}
\begin{barticle}[author]
\bauthor{\bsnm{Blei},~\bfnm{David~M}\binits{D.~M.}},
  \bauthor{\bsnm{Kucukelbir},~\bfnm{Alp}\binits{A.}} \AND
  \bauthor{\bsnm{McAuliffe},~\bfnm{Jon~D}\binits{J.~D.}}
(\byear{2017}).
\btitle{Variational inference: A review for statisticians}.
\bjournal{Journal of the American statistical Association}
\bvolume{112}
\bpages{859--877}.
\end{barticle}
\endbibitem

\bibitem[\protect\citeauthoryear{Celeux and
  Govaert}{1992}]{celeux1992classification}
\begin{barticle}[author]
\bauthor{\bsnm{Celeux},~\bfnm{Gilles}\binits{G.}} \AND
  \bauthor{\bsnm{Govaert},~\bfnm{G{\'e}rard}\binits{G.}}
(\byear{1992}).
\btitle{A classification EM algorithm for clustering and two stochastic
  versions}.
\bjournal{Computational statistics \& Data analysis}
\bvolume{14}
\bpages{315--332}.
\end{barticle}
\endbibitem

\bibitem[\protect\citeauthoryear{Chen and Li}{2013}]{chen2013variable}
\begin{barticle}[author]
\bauthor{\bsnm{Chen},~\bfnm{Jun}\binits{J.}} \AND
  \bauthor{\bsnm{Li},~\bfnm{Hongzhe}\binits{H.}}
(\byear{2013}).
\btitle{Variable selection for sparse Dirichlet-multinomial regression with an
  application to microbiome data analysis}.
\bjournal{The annals of applied statistics}
\bvolume{7}.
\end{barticle}
\endbibitem

\bibitem[\protect\citeauthoryear{Delyon, Lavielle and
  Moulines}{1999}]{delyon1999convergence}
\begin{barticle}[author]
\bauthor{\bsnm{Delyon},~\bfnm{Bernard}\binits{B.}},
  \bauthor{\bsnm{Lavielle},~\bfnm{Marc}\binits{M.}} \AND
  \bauthor{\bsnm{Moulines},~\bfnm{Eric}\binits{E.}}
(\byear{1999}).
\btitle{Convergence of a stochastic approximation version of the EM algorithm}.
\bjournal{Annals of statistics}
\bpages{94--128}.
\end{barticle}
\endbibitem

\bibitem[\protect\citeauthoryear{Duane et~al.}{1987}]{duane1987hybrid}
\begin{barticle}[author]
\bauthor{\bsnm{Duane},~\bfnm{Simon}\binits{S.}},
  \bauthor{\bsnm{Kennedy},~\bfnm{Anthony~D}\binits{A.~D.}},
  \bauthor{\bsnm{Pendleton},~\bfnm{Brian~J}\binits{B.~J.}} \AND
  \bauthor{\bsnm{Roweth},~\bfnm{Duncan}\binits{D.}}
(\byear{1987}).
\btitle{Hybrid monte carlo}.
\bjournal{Physics letters B}
\bvolume{195}
\bpages{216--222}.
\end{barticle}
\endbibitem

\bibitem[\protect\citeauthoryear{Gevers et~al.}{2014}]{gevers2014treatment}
\begin{barticle}[author]
\bauthor{\bsnm{Gevers},~\bfnm{Dirk}\binits{D.}},
  \bauthor{\bsnm{Kugathasan},~\bfnm{Subra}\binits{S.}},
  \bauthor{\bsnm{Denson},~\bfnm{Lee~A}\binits{L.~A.}},
  \bauthor{\bsnm{V{\'a}zquez-Baeza},~\bfnm{Yoshiki}\binits{Y.}},
  \bauthor{\bsnm{Van~Treuren},~\bfnm{Will}\binits{W.}},
  \bauthor{\bsnm{Ren},~\bfnm{Boyu}\binits{B.}},
  \bauthor{\bsnm{Schwager},~\bfnm{Emma}\binits{E.}},
  \bauthor{\bsnm{Knights},~\bfnm{Dan}\binits{D.}},
  \bauthor{\bsnm{Song},~\bfnm{Se~Jin}\binits{S.~J.}},
  \bauthor{\bsnm{Yassour},~\bfnm{Moran}\binits{M.}} \betal{et~al.}
(\byear{2014}).
\btitle{The treatment-naive microbiome in new-onset Crohn’s disease}.
\bjournal{Cell host \& microbe}
\bvolume{15}
\bpages{382--392}.
\end{barticle}
\endbibitem

\bibitem[\protect\citeauthoryear{Giordano, Broderick and
  Jordan}{2018}]{giordano2018covariances}
\begin{barticle}[author]
\bauthor{\bsnm{Giordano},~\bfnm{Ryan}\binits{R.}},
  \bauthor{\bsnm{Broderick},~\bfnm{Tamara}\binits{T.}} \AND
  \bauthor{\bsnm{Jordan},~\bfnm{Michael~I}\binits{M.~I.}}
(\byear{2018}).
\btitle{Covariances, robustness and variational bayes}.
\bjournal{Journal of machine learning research}
\bvolume{19}.
\end{barticle}
\endbibitem

\bibitem[\protect\citeauthoryear{MacKay}{2003}]{mackay2003information}
\begin{bbook}[author]
\bauthor{\bsnm{MacKay},~\bfnm{David~JC}\binits{D.~J.}}
(\byear{2003}).
\btitle{Information theory, inference and learning algorithms}.
\bpublisher{Cambridge university press}.
\end{bbook}
\endbibitem

\bibitem[\protect\citeauthoryear{Mosimann}{1962}]{mosimann1962compound}
\begin{barticle}[author]
\bauthor{\bsnm{Mosimann},~\bfnm{James~E}\binits{J.~E.}}
(\byear{1962}).
\btitle{On the compound multinomial distribution, the multivariate
  $\beta$-distribution, and correlations among proportions}.
\bjournal{Biometrika}
\bvolume{49}
\bpages{65--82}.
\end{barticle}
\endbibitem

\bibitem[\protect\citeauthoryear{Neal et~al.}{2011}]{neal2011mcmc}
\begin{barticle}[author]
\bauthor{\bsnm{Neal},~\bfnm{Radford~M}\binits{R.~M.}} \betal{et~al.}
(\byear{2011}).
\btitle{MCMC using Hamiltonian dynamics}.
\bjournal{Handbook of markov chain monte carlo}
\bvolume{2}
\bpages{2}.
\end{barticle}
\endbibitem

\bibitem[\protect\citeauthoryear{Opper and Saad}{2001}]{opper2001advanced}
\begin{bbook}[author]
\bauthor{\bsnm{Opper},~\bfnm{Manfred}\binits{M.}} \AND
  \bauthor{\bsnm{Saad},~\bfnm{David}\binits{D.}}
(\byear{2001}).
\btitle{Advanced mean field methods: Theory and practice}.
\bpublisher{MIT press}.
\end{bbook}
\endbibitem

\bibitem[\protect\citeauthoryear{Ormerod and
  Wand}{2010}]{ormerod2010explaining}
\begin{barticle}[author]
\bauthor{\bsnm{Ormerod},~\bfnm{John~T}\binits{J.~T.}} \AND
  \bauthor{\bsnm{Wand},~\bfnm{Matt~P}\binits{M.~P.}}
(\byear{2010}).
\btitle{Explaining variational approximations}.
\bjournal{The American Statistician}
\bvolume{64}
\bpages{140--153}.
\end{barticle}
\endbibitem

\bibitem[\protect\citeauthoryear{Seijas-Mac{\'\i}as
  et~al.}{2020}]{seijas2020approximating}
\begin{barticle}[author]
\bauthor{\bsnm{Seijas-Mac{\'\i}as},~\bfnm{Antonio}\binits{A.}},
  \bauthor{\bsnm{Oliveira},~\bfnm{Am{\'\i}lcar}\binits{A.}},
  \bauthor{\bsnm{Oliveira},~\bfnm{Teresa~A}\binits{T.~A.}} \AND
  \bauthor{\bsnm{Leiva},~\bfnm{V{\'\i}ctor}\binits{V.}}
(\byear{2020}).
\btitle{Approximating the distribution of the product of two normally
  distributed random variables}.
\bjournal{Symmetry}
\bvolume{12}
\bpages{1201}.
\end{barticle}
\endbibitem

\bibitem[\protect\citeauthoryear{Tang and Chen}{2019}]{tang2019zero}
\begin{barticle}[author]
\bauthor{\bsnm{Tang},~\bfnm{Zheng-Zheng}\binits{Z.-Z.}} \AND
  \bauthor{\bsnm{Chen},~\bfnm{Guanhua}\binits{G.}}
(\byear{2019}).
\btitle{Zero-inflated generalized Dirichlet multinomial regression model for
  microbiome compositional data analysis}.
\bjournal{Biostatistics}
\bvolume{20}
\bpages{698--713}.
\end{barticle}
\endbibitem

\bibitem[\protect\citeauthoryear{Tipping and
  Bishop}{1999}]{tipping1999probabilistic}
\begin{barticle}[author]
\bauthor{\bsnm{Tipping},~\bfnm{Michael~E}\binits{M.~E.}} \AND
  \bauthor{\bsnm{Bishop},~\bfnm{Christopher~M}\binits{C.~M.}}
(\byear{1999}).
\btitle{Probabilistic principal component analysis}.
\bjournal{Journal of the Royal Statistical Society Series B: Statistical
  Methodology}
\bvolume{61}
\bpages{611--622}.
\end{barticle}
\endbibitem

\bibitem[\protect\citeauthoryear{Wang and
  Titterington}{2005}]{wang2005inadequacy}
\begin{binproceedings}[author]
\bauthor{\bsnm{Wang},~\bfnm{Bo}\binits{B.}} \AND
  \bauthor{\bsnm{Titterington},~\bfnm{D~Michael}\binits{D.~M.}}
(\byear{2005}).
\btitle{Inadequacy of interval estimates corresponding to variational Bayesian
  approximations}.
In \bbooktitle{International Workshop on Artificial Intelligence and
  Statistics}
\bpages{373--380}.
\bpublisher{PMLR}.
\end{binproceedings}
\endbibitem

\bibitem[\protect\citeauthoryear{Wang and Zhao}{2017}]{wang2017dirichlet}
\begin{barticle}[author]
\bauthor{\bsnm{Wang},~\bfnm{Tao}\binits{T.}} \AND
  \bauthor{\bsnm{Zhao},~\bfnm{Hongyu}\binits{H.}}
(\byear{2017}).
\btitle{A Dirichlet-tree multinomial regression model for associating dietary
  nutrients with gut microorganisms}.
\bjournal{Biometrics}
\bvolume{73}
\bpages{792--801}.
\end{barticle}
\endbibitem

\bibitem[\protect\citeauthoryear{Weiss et~al.}{2016}]{weiss2016correlation}
\begin{barticle}[author]
\bauthor{\bsnm{Weiss},~\bfnm{Sophie}\binits{S.}},
  \bauthor{\bsnm{Van~Treuren},~\bfnm{Will}\binits{W.}},
  \bauthor{\bsnm{Lozupone},~\bfnm{Catherine}\binits{C.}},
  \bauthor{\bsnm{Faust},~\bfnm{Karoline}\binits{K.}},
  \bauthor{\bsnm{Friedman},~\bfnm{Jonathan}\binits{J.}},
  \bauthor{\bsnm{Deng},~\bfnm{Ye}\binits{Y.}},
  \bauthor{\bsnm{Xia},~\bfnm{Li~Charlie}\binits{L.~C.}},
  \bauthor{\bsnm{Xu},~\bfnm{Zhenjiang~Zech}\binits{Z.~Z.}},
  \bauthor{\bsnm{Ursell},~\bfnm{Luke}\binits{L.}},
  \bauthor{\bsnm{Alm},~\bfnm{Eric~J}\binits{E.~J.}} \betal{et~al.}
(\byear{2016}).
\btitle{Correlation detection strategies in microbial data sets vary widely in
  sensitivity and precision}.
\bjournal{The ISME journal}
\bvolume{10}
\bpages{1669--1681}.
\end{barticle}
\endbibitem

\bibitem[\protect\citeauthoryear{Xia et~al.}{2013}]{xia2013logistic}
\begin{barticle}[author]
\bauthor{\bsnm{Xia},~\bfnm{Fan}\binits{F.}},
  \bauthor{\bsnm{Chen},~\bfnm{Jun}\binits{J.}},
  \bauthor{\bsnm{Fung},~\bfnm{Wing~Kam}\binits{W.~K.}} \AND
  \bauthor{\bsnm{Li},~\bfnm{Hongzhe}\binits{H.}}
(\byear{2013}).
\btitle{A logistic normal multinomial regression model for microbiome
  compositional data analysis}.
\bjournal{Biometrics}
\bvolume{69}
\bpages{1053--1063}.
\end{barticle}
\endbibitem

\bibitem[\protect\citeauthoryear{Zeng et~al.}{2022}]{zeng2022zero}
\begin{barticle}[author]
\bauthor{\bsnm{Zeng},~\bfnm{Yanyan}\binits{Y.}},
  \bauthor{\bsnm{Pang},~\bfnm{Daolin}\binits{D.}},
  \bauthor{\bsnm{Zhao},~\bfnm{Hongyu}\binits{H.}} \AND
  \bauthor{\bsnm{Wang},~\bfnm{Tao}\binits{T.}}
(\byear{2022}).
\btitle{A zero-inflated logistic normal multinomial model for extracting
  microbial compositions}.
\bjournal{Journal of the American Statistical Association}
\bpages{1--14}.
\end{barticle}
\endbibitem

\bibitem[\protect\citeauthoryear{Zhang and Lin}{2019}]{zhang2019scalable}
\begin{barticle}[author]
\bauthor{\bsnm{Zhang},~\bfnm{Jingru}\binits{J.}} \AND
  \bauthor{\bsnm{Lin},~\bfnm{Wei}\binits{W.}}
(\byear{2019}).
\btitle{Scalable estimation and regularization for the logistic normal
  multinomial model}.
\bjournal{Biometrics}
\bvolume{75}
\bpages{1098--1108}.
\end{barticle}
\endbibitem

\end{thebibliography}

\end{document}